\PassOptionsToPackage{table}{xcolor}
\documentclass[sigconf, screen]{acmart}

\setcopyright{none}
\renewcommand\footnotetextcopyrightpermission[1]{} %
\makeatletter
\def\@affiliationfont{\vspace{5pt}\Large\itshape}
\g@addto@macro\@mkauthors{\vspace{6pt}} %
\makeatother
\usepackage{enumitem}

\usepackage[utf8]{inputenc}
\usepackage[english]{babel}
\usepackage{tikz}
\usetikzlibrary{decorations.pathreplacing,arrows.meta,patterns,calc, positioning, fit, backgrounds}
\usepackage{amsmath}
\usepackage{algorithm}
\usepackage{algpseudocode}

\algnewcommand\algorithmicinput{\textbf{Input:}}
\algnewcommand\algorithmicoutput{\textbf{Output:}}
\algnewcommand\Input{\item[\algorithmicinput]}
\algnewcommand\Output{\item[\algorithmicoutput]}

\usepackage{subcaption}
\usepackage{xspace}
\usepackage{url}
\usepackage{tabularx}
\newenvironment{packeditemize}{\begin{list}{$\bullet$}{\setlength{\itemsep}{0.1pt}\addtolength{\labelwidth}{0pt}\setlength{\leftmargin}{\labelwidth}\setlength{\listparindent}{\parindent}\setlength{\parsep}{1pt}\setlength{\topsep}{0pt}}}{\end{list}}
\providecommand{\alltoall}{\textsc{AllToAll}\xspace}
\providecommand{\allreduce}{\textsc{AllReduce}\xspace}
\providecommand{\allgather}{\textsc{AllGather}\xspace}

\providecommand{\reducescatter}{\textsc{ReduceScatter}\xspace}

\providecommand{\ie}{\emph{i.e.,} }
\providecommand{\eg}{\emph{e.g.,} }

\providecommand{\myparab}[1]{\vspace{0pt}\noindent\textbf{#1} }
\expandafter\def\expandafter\normalsize\expandafter{%
    \normalsize%
    \setlength\abovedisplayskip{1pt}%
    \setlength\belowdisplayskip{2pt}%
    \setlength\abovedisplayshortskip{0pt}%
    \setlength\belowdisplayshortskip{2pt}%
}
\newlist{compactitem}{itemize}{1}
\setlist[compactitem,1]{label=\textbullet, left=0pt, itemsep=1pt, topsep=1pt, parsep=0pt, partopsep=0pt}

\begin{document}

\setlength{\abovecaptionskip}{1pt plus 0pt minus 1pt}
\setlength{\belowcaptionskip}{-1pt}
\setlength{\textfloatsep}{5pt plus 1pt minus 1pt}
\setlength{\intextsep}{5pt plus 1pt minus 1pt}
\setlength{\floatsep}{5pt plus 1pt minus 1pt}
\setlength{\dbltextfloatsep}{4pt plus 1pt minus 1pt}
\setlength{\dblfloatsep}{4pt plus 1pt minus 1pt}

\title{The Cost of Synchronization in GPU Scale-Up Domains}

\author{\texorpdfstring{Arjun Devraj\qquad Lindsey Bowen\qquad Rachee Singh}{Arjun Devraj, Lindsey Bowen, Rachee Singh}}
\affiliation{%
  \institution{Cornell University}
  \country{}}
\begin{abstract}
GPU scale-up domains are the building blocks of machine learning infrastructure. Architects provision the size and bandwidth in these domains assuming all GPUs reach a collective operation at the same time. However, we find that this assumption does not hold in practice. Across 300,000 collectives from training four language models on A100, H100, and H200 domains, GPUs reach the same collective hundreds to thousands of microseconds apart, despite running identical kernels. Since a collective cannot complete until every GPU contributes its data, GPUs that arrive early wait for the slowest one. We call this waiting time the \emph{synchronization tax}, and it consumes nearly 40\% of tensor-parallel communication time in an 8-GPU domain. To find its source, we develop an algorithm that recovers the chain of operations that delays each GPU before a collective. Our algorithm attributes 78\% of the tax to variation in the runtimes of matrix-multiplication (GEMM) kernels across GPUs. Since the tax depends on the slowest of many GPUs, our analytical model shows that it grows at least logarithmically with domain size. Thus, a 72-GPU domain requires $3.3{\times}$ more per-GPU bandwidth than today's simulators predict to keep communication below 12\% of step time, and the optimal scale-up fabric should partition a switch's capacity among fewer GPUs with more bandwidth each.\footnote{Correspondence to \href{mailto:adevraj@cs.cornell.edu}{adevraj@cs.cornell.edu}}
\end{abstract}

\maketitle
\pagestyle{plain} %

\section{Introduction}
\label{sec:intro}

\begin{figure}[t]
\centering
\includegraphics[width=\linewidth]{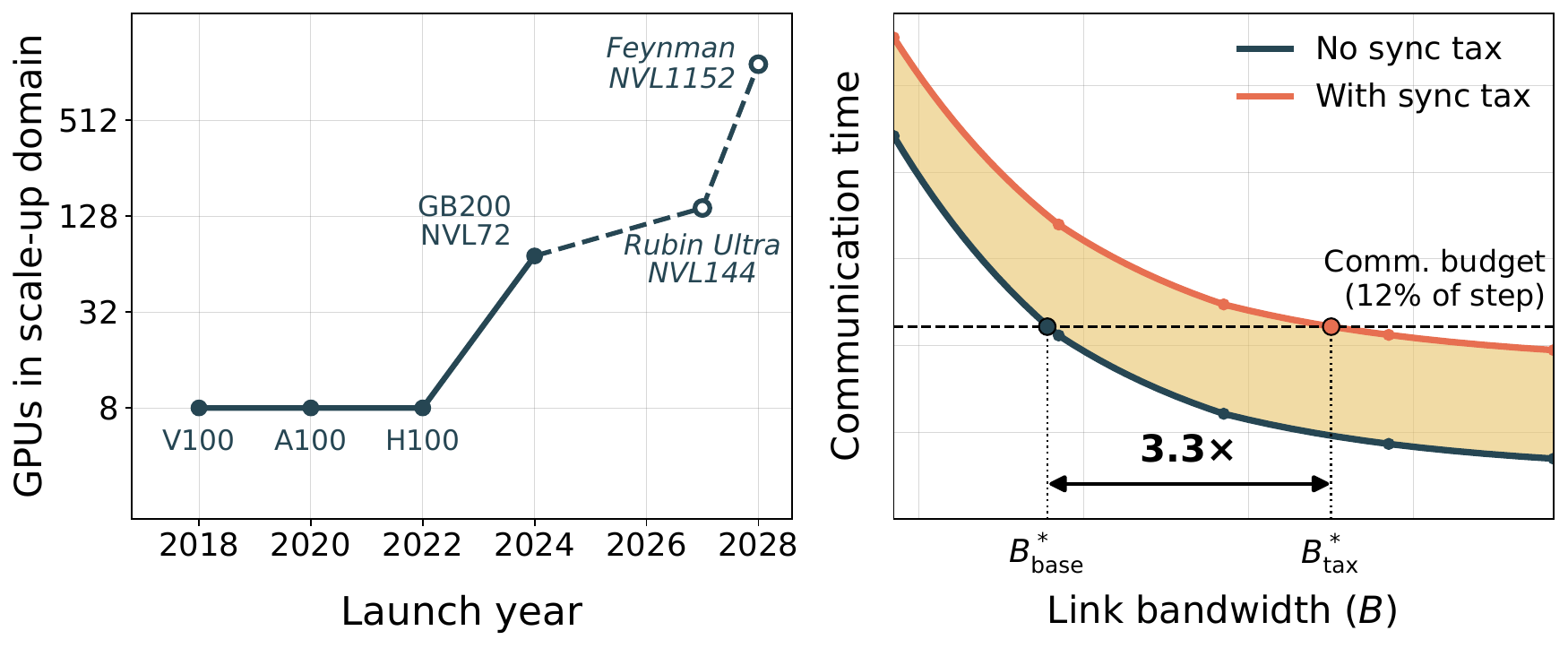}
\caption{GPU scale-up domains are growing from 8 to 1{,}152 GPUs (left). The synchronization tax (shaded) grows with domain size and imposes a communication overhead that bandwidth cannot reduce. After accounting for the tax, a 72-GPU domain requires $3.3{\times}$ more per-GPU bandwidth to guarantee that communication does not exceed 12\% of step time, \ie $B^{*}_{\text{tax}}$ vs.\ $B^{*}_{\text{base}}$ (right).}
\label{fig:intro}
\end{figure}

Scale-up domains like multi-GPU servers (\eg NVIDIA DGX) or racks (\eg NVL72) are the building blocks of machine learning (ML) infrastructure. Within a scale-up domain, GPUs communicate over a uniform, high-bandwidth fabric. The datacenter network then connects many such domains into ML datacenters the size of a warehouse~\cite{qian2024alibaba,jiang2024megascale,megatron,rasley2020deepspeed,hu2024characterization}. In the past five years, GPU manufacturers have expanded both the size and inter-GPU bandwidth of scale-up domains, tripling per-GPU interconnect bandwidth from 300 to 900~GB/s and increasing the domain size from 8 to 72 GPUs (Figure~\ref{fig:intro}, left). Roadmaps now extend scale-up domains to 1{,}152 GPUs and terabytes per second of per-GPU bandwidth~\cite{zuo2025serving,nvidia-vera-rubin-pod-2026,nvl1152}.

This explosion in the size and inter-GPU bandwidth of scale-up domains has been driven by the growing scale of ML workloads, which require partitioning computation across many GPUs. A typical ML workload executes as a sequence of compute phases separated by collective communication (\eg \allreduce). For instance, \emph{tensor parallelism} (TP) shards each transformer layer's general matrix multiplications (GEMMs) across the domain, so all participating GPUs, or \emph{ranks}, exchange data in every layer. Each collective, regardless of its implementation, requires every rank to contribute its data, thereby acting as a synchronization point at which ranks must wait for the slowest one among them~\cite{devraj2026breaking}.

Since every model layer must wait for one or more collectives to complete, scale-up domains are designed to minimize the overheads of frequent collective communication. To design future domains, architects rely on analytical models (\eg Hockney, or $\alpha{-}\beta$ model~\cite{alphabeta}) and multi-GPU simulators~\cite{won2023astra,khairy2020accel} to determine the number of GPUs in the domain and how much bandwidth to provision between them. However, these tools share a common assumption that all GPUs arrive at the collective simultaneously, so communication time is dictated by latency and bandwidth alone.

In this work, we show that this assumption falls short in practice. We profile the fine-tuning of four recent LLMs (Llama-3 8B~\cite{grattafiori2024llama}, Qwen-3 32B~\cite{yang2025qwen3}, DeepSeek-V3 16B~\cite{liu2024deepseek}, Llama-3 70B~\cite{grattafiori2024llama}) on scale-up domains of A100, H100, and H200 GPUs. Across the ${\sim}$300,000 collectives in these workloads, ranks arrive at the same collective hundreds to thousands of microseconds apart, even though they execute identical kernels on architecturally identical hardware over a uniform fabric. In an 8-GPU domain, ranks can spend nearly 40\% of their TP communication time waiting for the slowest rank to arrive (\S\ref{sec:characterization}). The identity of the slowest rank changes from one collective to the next, which rules out a single faulty device as the cause~\cite{lin2025understanding,jiang2024megascale,wu2024falcon}. As a result, communication time is dictated not only by the latency and bandwidth of the network, but also by this unproductive wait. We refer to this wait as the \emph{synchronization tax} because it taxes the gains from scaling up ML infrastructure.

Whether architects can eliminate the tax or must provision for it depends on why ranks arrive at different times. While prior work has explored runtime variability in GPUs~\cite{go2025characterizing,kurzynski2025lit,sinha2022not,chung2025swift}, a slower or higher-variance kernel is not necessarily responsible for the synchronization tax. Many kernels execute concurrently across CUDA streams on the GPU, and many may not lie on the \emph{critical path}, \ie the chain of events that determines when a rank arrives at the collective. Identifying this critical path is challenging because the CPU dispatches GPU kernels asynchronously, and a proprietary GPU scheduler executes each kernel once its dependencies complete. The dependency that ultimately delays the slowest rank's arrival at a collective is resolved only at runtime, and any of the thousands of preceding kernels is a potential culprit. We find that existing trace analysis tools cannot resolve this challenge~\cite{bhatnagar2022hta, lin2025understanding,guan2025perftracker} because they target \emph{severe slowdowns}, not the routine microsecond-scale variation that produces the synchronization tax.

Our key insight is that all ranks finish the previous collective at nearly the same time, so any skew at the current collective must originate from events between the two. This window bounds the search and gives all ranks a common reference point, even though their traces do not share a clock. We build a dependency graph over the CPU and GPU events in this window and trace backward from each rank's arrival through its critical path. Our algorithm can attribute runtime synchronization delays at \emph{sub-$10\,\mu$s granularity} to specific GPU kernels. We validate that our algorithm (1) recovers delays that we inject artificially and (2) predicts the synchronization delays we observe across hundreds of thousands of collectives spanning GPU architectures \emph{solely} from preceding GPU kernels and CPU events in the trace, unlike common heuristics such as selecting the longest-running kernel (\S\ref{sec:attribution}).

Our algorithm attributes 78\% of the tax to cross-rank variation in GEMM kernel runtimes, despite equal work on every rank. We rule out input-dependent causes (sparsity, tensor shapes) and identify dynamic clock boosting on GPUs as a potential culprit. Although pinning GPU clocks reduces the tax, it does not eliminate it and slows down training by up to 14\% (\S\ref{sec:gemm-variation}). As a result, this variation cannot be cheaply engineered away, and the tax will persist in future scale-up domains.

Architects must provision for the tax at scales they cannot measure, but hardware simulators assume deterministic kernel runtimes. Thus, we develop an analytical model of the synchronization tax using extreme value theory (EVT). Our attribution results allow us to ignore one-off noise and model the dominant source of the tax: routine GEMM variation. Our model predicts the expected tax for an $n$-GPU scale-up domain solely using the mean and variance of key GEMM kernels collected from a small-scale ($\ll n$) sample. Even conservative estimates show that the tax grows \emph{at least logarithmically} with domain size. We validate that our model accurately predicts the observed synchronization tax on held-out workloads and show that even when fit on a 4-GPU run, it can predict the variation we measure in a 128-GPU job (\S\ref{sec:tax-model}).

We incorporate our analytical model of the tax into ASTRA-sim~\cite{won2023astra}, a state-of-the-art hardware simulator for distributed ML, and find that it changes three key provisioning decisions for future scale-up domains. First, we show that bounding communication at 12\% of step time on a 72-GPU scale-up domain (\ie NVIDIA's GB200 NVL72) requires $3.3{\times}$ more interconnect bandwidth than predicted by today's tax-free simulators (Figure~\ref{fig:intro}, right) because the synchronization tax imposes a bandwidth-independent overhead. Since the tax grows with the number of GPUs, expanding domain size from 72 to 144 GPUs adds as much communication time as doubling per-GPU bandwidth from 3.6 to 7.2~TB/s removes. Second, we explore how to optimally partition a scale-up switch's capacity among GPUs in the domain, navigating the trade-off between connecting more GPUs with less bandwidth each vs.\ fewer GPUs with more bandwidth each. After incorporating the tax, the optimum moves toward fewer GPUs with more bandwidth each (up to 42\% fewer GPUs with $1.7{\times}$ bandwidth). In fact, today's simulators underestimate the communication time of their optimal designs by $1.5$--$3{\times}$. Finally, we show that if reductions in GEMM variation continue to lag behind compute speedups, today's simulators will underestimate communication's share of end-to-end time by $2.1{\times}$.

In summary, we make the following contributions:
\begin{compactitem}
\item A measurement study of the synchronization tax, collected from kernel-level traces that include 300K collectives spanning four ML workloads and three generations of GPU scale-up domains.
\item An attribution algorithm that identifies critical paths \emph{across ranks} for each collective, pinning 78\% of the tax at the collective on cross-rank variation in GEMM execution times.
\item An analytical model that can predict the tax for an $n$-GPU domain solely from GEMM statistics. We validate that our model predicts the empirical tax across measured workloads, and that it captures GEMM variation measured on a 128-GPU job.
\item An integration of our analytical model into hardware simulators that shows that the tax changes key architectural provisioning decisions, like increasing the per-GPU interconnect bandwidth to offset the tax and moving switched scale-up designs toward fewer GPUs with more bandwidth each relative to baselines.
\end{compactitem}

\section{Background}
\label{sec:background}
In this section, we introduce GPU scale-up domains and discuss how distributed ML workloads execute on these domains.

\subsection{Scale-Up Domain Architecture}
\label{sec:bg-arch}

A \emph{scale-up domain} is a set of GPUs interconnected by a high-bandwidth fabric within a physical enclosure, typically a multi-GPU server or, in latest generations, a full rack:

\vspace{1mm}
{\centering
\definecolor{cpuBlue}{RGB}{38,70,83}
\definecolor{gpuGreen}{RGB}{42,157,143}
\definecolor{commPurple}{RGB}{244,162,97}
\begin{tikzpicture}[font=\footnotesize,
  gpu/.style={fill=gpuGreen!80,draw=gpuGreen!55!black,line width=0.5pt,text=black,minimum width=0.72cm,minimum height=0.38cm,inner sep=0pt,font=\scriptsize},
  bar/.style={fill=black!30,draw=black!65,line width=0.5pt,text=black,minimum height=0.36cm,inner sep=0.5pt},
  cpu/.style={fill=cpuBlue!55,draw=cpuBlue!80!black,line width=0.5pt,text=black,minimum height=0.36cm,inner sep=0.5pt},
  nic/.style={fill=commPurple!80,draw=commPurple!60!black,line width=0.5pt,text=black,minimum height=0.36cm,inner sep=0.5pt},
  nvlink/.style={commPurple,line width=1.2pt},
  pcie/.style={cpuBlue!80,line width=0.8pt,densely dashed}]
\foreach \i in {0,...,7}{ \node[gpu] (g\i) at (0.36+0.8*\i,0) {GPU\,\i}; }
\node[bar,minimum width=6.32cm] (nvs) at ($(g0)!0.5!(g7)+(0,0.62)$) {Non-blocking switch (\eg NVSwitch)};
\foreach \i in {0,...,7}{ \draw[nvlink] (g\i.north) -- (g\i.north |- nvs.south); }
\node[cpu,minimum width=1.52cm,anchor=west] (cpu) at ($(g0.west)+(0,-0.6)$) {Host CPU};
\node[nic,minimum width=1.52cm,anchor=east] (nic) at ($(g7.east |- cpu)$) {Scale-out NIC};
\draw[pcie] (cpu.east) -- (nic.west);
\foreach \i in {0,1}{ \draw[pcie] (g\i.south) -- (g\i.south |- cpu.north); }
\foreach \i in {6,7}{ \draw[pcie] (g\i.south) -- (g\i.south |- nic.north); }
\foreach \i in {2,...,5}{ \draw[pcie] (g\i.south) -- (g\i.south |- cpu.east); }
\node[fill=white,inner sep=1pt,font=\scriptsize] at ($(g3.south |- cpu.east)!0.5!(g4.south |- cpu.east)$) {PCIe};
\node[draw=black!45,line width=0.5pt,inner xsep=3pt,inner ysep=3pt,fit=(nvs)(g0)(cpu)(nic)] (box) {};
\node[fill=white,inner sep=1pt,anchor=west,font=\scriptsize\itshape] at ([xshift=6pt]box.north west) {Scale-up domain (server or rack)};
\node[anchor=west,align=left,font=\scriptsize,inner sep=0pt] at ($(box.east |- g7.north)!0.5!(box.east |- nvs.south)+(3pt,0)$) {NVLink\\300--900\,GB/s\\per GPU};
\draw[-{Stealth[length=3pt]},commPurple,line width=0.8pt] (nic.east) -- ($(box.east |- nic.east)+(8pt,0)$)
  node[anchor=west,align=left,font=\scriptsize,text=black,inner sep=1pt] {other\\domains};
\end{tikzpicture}
\par}
\vspace{1mm}

\myparab{Interconnect.}
GPUs in a scale-up domain communicate over interconnects (\eg NVIDIA's NVLink and NVSwitch~\cite{nvlink}, AMD's UALink~\cite{ualink}) that offer an order of magnitude higher bandwidth than the datacenter network connecting these domains. NVLink provides bidirectional links between GPUs, and NVSwitch~\cite{nvlink, ishii2022nvlink} connects them through a non-blocking fabric. Per-GPU NVLink bandwidth has grown from 300~GB/s on A100 to 900~GB/s on B200, and domain size has expanded from 8 to 72 GPUs, with plans for 1,152 GPUs and 10 PB/s of aggregate bandwidth~\cite{nvidia-vera-rubin-pod-2026}.

\myparab{Topology.} Scale-up domains provide a fully connected topology where every GPU pair has a direct path with uniform latency and bandwidth~\cite{nvl72}. Similarly, Huawei's CloudMatrix connects 384 accelerators with an all-to-all network~\cite{zuo2025serving}. In contrast, \emph{scale-out} networks (\eg InfiniBand~\cite{infiniband}, RoCE~\cite{roce}) connect scale-up domains using multi-hop topologies with links shared across flows, so the latency and bandwidth between two GPUs vary with the path they take and with competing traffic on the scale-out network.

\subsection{Distributed ML in Scale-Up Domains}
\label{sec:bg-workloads}

LLM training and inference distribute computation across GPUs using various \emph{parallelism} strategies. Within a scale-up domain, the common parallelism strategies are \emph{tensor parallelism} (TP) and \emph{fully sharded data parallelism} (FSDP), which are often composed together into hybrid parallelism~\cite{narayanan2021efficient, megatron}. Mixture-of-experts (MoE) models also use \emph{expert parallelism} (EP) to shard expert layers across GPUs, which incurs \alltoall collectives in every MoE layer~\cite{rajbhandari2022deepspeed, hwang2023tutel}.

\myparab{TP} partitions large matrix multiplications across GPUs~\cite{narayanan2021efficient}. Each GPU computes a partial matrix multiply, and the partial results are combined through an \allreduce collective (equivalently, \reducescatter + \allgather) in each transformer sublayer. Since collective communication in TP is ``exposed,'' \ie not overlapped with computation, and uses large data buffers, TP is often restricted to high-bandwidth scale-up domains.

\myparab{FSDP} shards model parameters, gradients, and optimizer state across GPUs~\cite{zhao2023pytorch}. Before each layer's forward pass, an \allgather reconstructs the full tensor from its shards. After the backward pass, a \reducescatter distributes and reduces gradients. FSDP is frequently composed with TP in hybrid parallelism.

\myparab{Host-driven execution model.} A training iteration executes as a series of GPU \emph{kernels} for both compute (\eg GEMMs) and collective communication (\eg \allreduce). The host CPU dispatches kernels asynchronously onto one or more CUDA streams per GPU, without waiting for any kernel to complete before issuing the next. The GPU scheduler executes each kernel once its dependencies (\eg other kernels) are complete~\cite{nsys}. Thus, the start of the kernel is determined by the joint state of the CPU dispatch queue and the GPU's internal scheduling state at runtime. Each GPU is typically managed by a separate CPU process~\cite{torchtitan}.

\myparab{Collectives are implicit synchronization points.} During a collective, a group of GPUs exchanges partial results. For example, each element of an \allreduce output is a sum over the inputs of all ranks. As a result, a collective cannot complete on \emph{any} rank without \emph{every} rank communicating its portion of the data~\cite{devraj2026breaking}. While earlier implementations placed a global barrier before the collective to enforce this dependency, modern collective communication libraries (CCLs) like NVIDIA's NCCL instead opt for lazy, flag-based synchronization~\cite{hu2025demystifying}. This enables each rank to start the collective independently, once its own preceding kernels finish. However, if one rank finishes its previous work later than the others, the ranks that start earlier wait inside the collective for its data because of the data dependencies imposed by the schedule to guarantee a correct result~\cite{devraj2026breaking}. We call this wait the \emph{synchronization delay} (\S\ref{sec:bg-synctax}).

\subsection{The Synchronization Delay}
\label{sec:bg-synctax}
CCLs implement collectives using parallel algorithms that specify data-transfer schedules. These algorithms are designed under the Hockney ($\alpha{-}\beta$) cost model~\cite{alphabeta, taccl,won2023astra,wang2025simai}, which estimates the time to send $S$ bytes as $\alpha + S{\cdot}{\beta}$, where $\alpha$ is the fixed startup latency of a message and $\beta{=}1{/}B$ with link bandwidth $B$. The time to complete a collective is $T{=}p{\cdot}\alpha + q{\cdot}S{\cdot}\beta$, where $p$ and $q$ are the coefficients on the latency and bandwidth costs set by the collective algorithm.

\myparab{Synchronization delay.} The Hockney model only measures communication time \emph{once all ranks are ready}. However, a collective requires all ranks to participate in the transfer schedule to correctly include every rank's data. Thus, data transfers within a collective communication schedule can be delayed due to \emph{synchronization}, \ie waiting on the slowest rank to provide its data. We formally define the synchronization delay for rank $i \in \{1,{\ldots},n\}$ as:

\noindent\refstepcounter{equation}\label{eq:synctax}\hfill
$\vcenter{\hbox{{\setlength{\fboxsep}{4pt}\colorbox{gray!12}{Eqn~\theequation:\enspace$\displaystyle \tau_i = \max_{j}\, t_j^{\text{start}} - t_i^{\text{start}}$}}}}$%
\hfill
$\vcenter{\hbox{\begin{tikzpicture}[x=0.24cm,y=0.33cm,font=\small]
\definecolor{gpuGreen}{RGB}{42,157,143}
\definecolor{commPurple}{RGB}{244,162,97}
\tikzset{comp/.style={draw=gpuGreen!80,fill=gpuGreen!25,line width=0.4pt},
         xfer/.style={draw=commPurple!90!black,fill=commPurple!60,line width=0.4pt},
         wait/.style={draw=commPurple!90!black,pattern=north east lines,pattern color=commPurple!90!black,line width=0.4pt}}
\foreach \r/\s/\y in {1/3/2, 2/4.5/1, 3/6/0} {
  \node[anchor=east,inner sep=1pt] at (-0.2,\y+0.25) {rank \r};
  \draw[comp] (0,\y) rectangle (\s,\y+0.5);
}
\draw[wait] (3,2) rectangle (6,2.5);
\draw[wait] (4.5,1) rectangle (6,1.5);
\foreach \y in {0,1,2} \draw[xfer] (6,\y) rectangle (9.5,\y+0.5);
\node[fill=white,inner sep=0.3pt] at (4.5,2.25) {$\tau_1$};
\draw[gray] (3,2.5) -- (3,2.85);
\node[above,inner sep=1pt] at (3,2.8) {$t_1^{\text{start}}$};
\draw[densely dashed,gray] (6,-0.3) -- (6,2.8);
\draw[densely dashed,gray] (9.5,-0.3) -- (9.5,2.8);
\node[below,inner sep=1pt] at (6,-0.3) {$\max_j t_j^{\text{start}}$};
\end{tikzpicture}}}$%
\hfill\null\par
\vspace{-1mm}

where $t_i^{\text{start}}$ is the time when rank $i$ enters the collective. We can define $\tau_i$ by start times alone because the remaining transfers take nearly the same time on every rank once the last rank arrives since the scale-up fabric is uniform~\cite{devraj2026breaking}. Ranks therefore start a collective at different times but finish it at nearly the same time, so $\tau_i$ is the time rank $i$ idles inside the collective. Unless otherwise specified, $\tau$ denotes the \emph{expected} (\ie mean) synchronization delay across ranks. A key property is that the synchronization delay is independent of interconnect bandwidth. While bandwidth impacts the transfer phase once all ranks are ready, it cannot reduce the idle time during the collective waiting on a slower rank.

\section{Measuring the Synchronization Delay}
\label{sec:characterization}

\begin{figure*}[t]
\centering
\includegraphics[width=\linewidth]{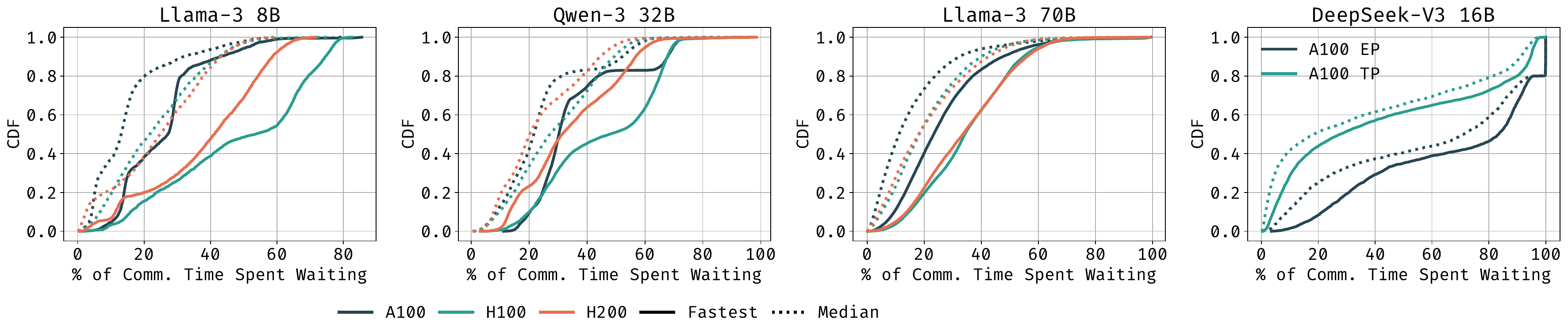}
\caption{CDFs of the synchronization tax---shown as the percent of NCCL collective communication time that the fastest and median ranks spend waiting---across TP collectives (\reducescatter) for all workloads and architectures (with EP for DeepSeek).}
\label{fig:synctax-cdf}
\end{figure*}

We measure the synchronization tax across ML jobs and scale-up domains, and find that it constrains performance.

\subsection{Experimental Setup}
\label{sec:setup}

We profile scale-up domains built from three recent NVIDIA GPU models (A100, H100, H200). All GPUs are the same model in each domain. The A100 experiments run on NERSC's Perlmutter supercomputer~\cite{nersc_perlmutter_architecture}, where each node has four 40GB-A100 GPUs with 100 GB/s peer-to-peer (P2P) NVLink bandwidth, and on commercial cloud instances~\cite{runpod2025} with 4--8 80GB-A100 GPUs per node and 300 GB/s P2P bandwidth via NVLink/NVSwitch~\cite{nvlink}. The H100/H200 experiments run on 8-GPU cloud instances with 450 GB/s P2P bandwidth via NVLink/NVSwitch.

\myparab{ML models and configurations.} We study four LLM workloads: Llama-3 8B~\cite{grattafiori2024llama}, DeepSeek-V3 16B~\cite{liu2024deepseek}, Qwen-3 32B~\cite{yang2025qwen3}, and Llama-3 70B~\cite{grattafiori2024llama}. All experiments perform supervised fine-tuning using the \texttt{torchtitan} framework~\cite{torchtitan}. We use local batch sizes of 4 for Llama, 8 for Qwen, and 2 for DeepSeek-V3 (since it runs on the smaller 40GB-A100 Perlmutter nodes), and sequence lengths of 4096 for Llama/DeepSeek and 8192 for Qwen. These combinations saturate GPU memory, reflecting realistic training setups.

Each workload in our experiments uses TP. Llama-3 8B runs 4-way TP on four GPUs and Qwen-3 32B runs 8-way TP on eight GPUs, so in both workloads a single TP group spans all GPUs in the job. Llama-3 70B and DeepSeek-V3 run on eight GPUs and combine 4-way TP with 2-way FSDP. This hybrid configuration divides the eight GPUs into two TP groups of four GPUs and four FSDP groups of two GPUs, and each GPU belongs to one group of each kind. DeepSeek-V3 is a mixture of experts (MoE) model, so we also evaluate expert parallelism (EP), which shards experts across GPUs and incurs frequent \alltoall communication. We note that unlike TP, which evenly partitions work across ranks, EP can incur load imbalance~\cite{ren2025enabling,li2023accelerating,rajbhandari2022deepspeed}. Table~\ref{tab:workloads} summarizes our workloads.

\begin{table}[h]
\centering
\small
\resizebox{\columnwidth}{!}{
\begin{tabular}{lccccccc}
\toprule
Model & Params & Precision & GPUs & TP & FSDP & EP & Architectures \\
\midrule
Llama-3 & 8B & FP32 & 4 & 4 & -- & -- & A100, H100, H200 \\
DeepSeek-V3 & 16B & BF16 & 8 & 4 & 2 & (4,8) & A100 (40 GB HBM) \\
Qwen-3 & 32B & FP32 & 8 & 8 & -- & -- & A100, H100, H200 \\
Llama-3 & 70B & BF16 & 8 & 4 & 2 & -- & A100, H100, H200 \\
\bottomrule
\end{tabular}
}
\caption{Workload configurations.}
\label{tab:workloads}
\vspace{-1em}
\end{table}

\subsection{Profiling Methodology}
\label{sec:profiling}
We profile training iterations using PyTorch's Kineto profiler~\cite{pytorch_profiler}, which uses the CUDA Profiling Tools Interface (CUPTI) to record timestamped GPU kernel executions, CUDA API calls, CUDA driver and runtime events (\eg kernel launches, \texttt{cudaStreamSynchronize}), and CPU-side functions, at microsecond ($\mu s$) resolution. Kineto offers low-overhead profiling which is critical for collecting frequent, per-rank traces without distorting runtime behavior. Each rank produces a JSON trace with per-event records (\eg timestamp, duration, CUDA stream). We collect traces every 5--10 iterations for a total of 50--100 iterations. Each workload generates several GB of trace data. We analyze roughly 293K collective operations in total.

Computing the synchronization delay in Eq.~\ref{eq:synctax} from the collected traces requires us to identify the time at which each rank enters the same collective. However, the workload traces fall short at this task for several reasons. First, each rank produces its own trace and the traces do not carry an identifier that links a collective on one rank to the same collective on another. Second, each trace interleaves the collectives of every parallelism group that the rank belongs to (\eg TP and FSDP) making it difficult to isolate the synchronization delay per group. Third, timestamps are not comparable across traces from different ranks as they do not share a common clock, so we cannot subtract the entry time on one rank from the entry time on another. We resolve these challenges with the following post-processing steps, which we validate in \S\ref{sec:attribution}:

\begin{compactitem}

\item \myparab{Separating process groups.} PyTorch assigns each parallelism (TP, FSDP, EP) a \emph{process group}, \ie the set of ranks that communicate with each other, and issues each group's collectives on a dedicated CUDA stream. We use the CUDA stream recorded in the traces to identify the process group of each NCCL kernel.

\item \myparab{Matching collectives across ranks.} A CUDA stream executes kernels in the order in which they were issued. Moreover, all ranks in a process group issue the same sequence of collectives. Consequently, the $k$-th NCCL kernel on the group's stream corresponds to the same collective on every rank.

\item \myparab{Computing synchronization delay.} We cannot compare kernel start timestamps across ranks because they do not share the same clock. Instead, we compare kernel durations, which cancel out fixed clock offsets between ranks. Recall that a collective does not complete on any rank until the last rank enters the collective since all ranks must contribute their data (\S\ref{sec:bg-workloads}). From that point, the transfer takes nearly the same time on every rank since the scale-up fabric has uniform bandwidth and each rank exchanges the same volume of data. All ranks therefore finish a collective at nearly the same time, and the NCCL kernel duration $d_i$ on rank $i$ equals this common end time minus the time at which rank $i$ entered the collective. Eq.~\ref{eq:synctax} then reduces to $\tau_i = d_i - \min_j d_j$, which uses only durations measured within a trace. We report two delays for each collective, that of the rank that enters first ($\max{-}\min$) and that of the median rank ($\text{median}{-}\min$), and validate the common end time assumption in \S\ref{sec:attribution} (Figure~\ref{fig:synctax-scatter}).

\end{compactitem}

\subsection{Severity of the Synchronization Delay}
\label{sec:prevalence}
Figure~\ref{fig:synctax-cdf} shows the measured synchronization delay across all TP \reducescatter collectives and EP \alltoall{'s} for every workload and GPU model in our study. The delay is a significant concern across all 10 configurations. Even in an 8-GPU scale-up domain, the median-arriving rank can spend over 80\% of its communication time waiting on the slowest rank to start the collective. This trend \emph{worsens} with hardware generation. For the largest model (Llama-3 70B) and newest GPU architecture (H200) measured, the fastest rank spends a median of 40\% and, at worst, nearly \emph{all} of its \reducescatter time waiting. For DeepSeek, the median rank spends over 60\% of its EP communication time waiting. We provide more exhaustive per-collective and FSDP results in Appendix~\ref{sec:appendix-measurements}.

Our findings contradict popular intuition that a scale-up domain behaves as a single large accelerator. GPUs in a scale-up domain are architecturally identical, share a uniform high-bandwidth network fabric, and complete identical work under TP, yet routinely arrive at collectives hundreds to thousands of microseconds apart. Moreover, we find that these delays are not caused by a single device, as the identity of the slowest rank changes over time (Appendix~\ref{sec:slowest-id}). Our findings suggest the synchronization delay is \emph{intrinsic} to distributed computation, rendering typical detection and mitigation (\eg job restart, device replacement) ineffective.

\section{Tracing the Synchronization Delay}
\label{sec:attribution}
We study why ranks enter the same collective at different times. A rank can begin the collective when two conditions are met: (1) its CPU has launched the kernel and (2) GPU kernels that produce its input have finished execution. A delay in either of these events can delay the rank's entry into the collective. However, each of these events in turn can be delayed by earlier events, and so on. Therefore, we need to trace back through the chain of events that precede a collective to understand why ranks enter the collective at different times. We call this chain of events the rank's \emph{critical path} for the collective. By definition, each event on the critical path is the last of its successor's dependencies to finish, so an increase in duration of an event on the critical path will delay the rank's entry into the collective. We develop an algorithm to recover critical paths of each rank for every collective, and use it to attribute synchronization delay to events on ranks.

\myparab{Why is this challenging?} Recovering critical paths from traces is challenging for two reasons. First, a trace simply records when each kernel executed and lacks information about dependencies that prevented the kernel from starting. Because the CPU launches kernels asynchronously, a kernel may start long after its launch, only after previous GPU work finishes. Further, kernels on different CUDA streams run concurrently. A kernel on one stream may wait for a kernel on another stream (\eg GEMM waiting on an \allgather that produces its weights) without the trace explicitly recording it. Second, each rank's trace contains thousands of CPU and GPU events per iteration, so a brute-force search is infeasible.

\myparab{Limitations of existing tools.} PyTorch's Holistic Trace Analysis (HTA)~\cite{bhatnagar2022hta} computes the critical path of a region within a single rank's trace but it requires enabling explicit CUDA synchronization events  to recover cross-stream dependencies. These events alter runtime behavior, making it challenging to apply the findings to the original workload. EROICA~\cite{guan2025perftracker} selects events by a fixed priority among event types instead of following dependencies, while PRISM~\cite{golden2025prism} and Lumos~\cite{liang2025lumos} model at iteration, rather than per collective, granularity. Appendix~\ref{sec:eroica} provides details on why existing tools are insufficient. Thus, we design a novel algorithm to recover the critical path on each rank before a collective. Our algorithm (Figure~\ref{fig:algo}) builds a dependency graph over the events of each rank (\S\ref{sec:dep-graph}), walks backward from a collective to extract and align the critical paths of all ranks (\S\ref{sec:critical-path}), and identifies the events that make the critical path of the last rank to enter longer than others.

\subsection{Dependency Graph Construction}
\label{sec:dep-graph}
In each training iteration, we build a directed graph for every rank in which the nodes are GPU kernels or CPU operations. An edge from event $a$ to event $b$ indicates that $b$ may depend on $a$ (Figure~\ref{fig:algo}).
\myparab{Nodes.} Each node stores the event's name, start time, duration, and CUDA stream or CPU thread. Kineto records the CPU operations of each thread as nested intervals, \ie each event's call stack. To avoid double-counting time on CPU, we split each CPU event's interval at the start and end times of its nested events, so each CPU event only stores the time during which none of its nested events run.

\myparab{Recorded dependencies.} We add an edge from each event to the next event on the same GPU stream or CPU thread, from each CPU kernel launch to the kernel it launches (matched by CUPTI correlation identifier), and from the most recently completed GPU kernel to each CPU \texttt{cudaStreamSynchronize} call (Appendix~\ref{sec:algo}).

\myparab{Inferred dependencies.} Traces do not record when a kernel on one stream waits for a kernel on another, so we infer these cross-stream dependencies from timing. We process events in order of start time and keep in-flight events in a min-heap keyed by completion time. When an event starts at time $t$, we pop every entry that completed at or before $t$ and add an edge from the last entry popped, \ie the event on any stream that completed most recently before $t$, since it is the last event that could have delayed the new one. We then push the new event onto the heap.

\begin{figure}
\centering
\usetikzlibrary{arrows.meta, positioning, fit, backgrounds, calc, decorations.pathreplacing}

\definecolor{cpuBlue}     {RGB}{38,70,83}
\definecolor{gpuGreen}    {RGB}{42,157,143}
\definecolor{commPurple}  {RGB}{244,162,97}
\definecolor{commOrange}  {RGB}{231,111,81}
\definecolor{traceBg}     {RGB}{245,245,250}
\definecolor{headerBg}    {RGB}{38,70,83}
\definecolor{pathColor}{RGB}{220,20,60}

\resizebox{0.84\columnwidth}{!}{%
\begin{tikzpicture}[scale=1.0, transform shape,
  causal/.style  = {-{Stealth[length=4pt]}, line width=1.5pt, pathColor!80!black},
  dagedge/.style = {-{Stealth[length=4pt]}, black!70, line width=0.9pt},
  font=\footnotesize\ttfamily,
  cpuC/.style  = {circle, draw=cpuBlue,    fill=cpuBlue!15,
                line width=1.0pt, minimum size=0.46cm, font=\normalsize,
                text=cpuBlue,    inner sep=1pt},
gpuC/.style  = {circle, draw=gpuGreen,   fill=gpuGreen!15,
                line width=1.0pt, minimum size=0.46cm, font=\normalsize,
                text=gpuGreen!60!black,  inner sep=1pt},
commC/.style = {circle, draw=commPurple!80!black, fill=commPurple!15,
                line width=1.0pt, minimum size=0.46cm, font=\normalsize,
                text=commPurple!50!black,inner sep=1pt},
commOC/.style= {circle, draw=commOrange, fill=commOrange!15,
                line width=1.6pt, minimum size=0.46cm, font=\normalsize,
                text=commOrange!50!black,inner sep=1pt},
]

\def\TW{6.5}
\def\RH{0.44}
\def\LW{-.6}
\def\TX{-1.8}

\def\BraceX{-0.12}   %
\def\LabelX{-0.55}   %

\def\R0Y{0}
\def\Rz{\R0Y}

\foreach \ry in {0,1,2,3}{
  \draw[gray!40,very thin] (\TX+\LW,\Rz+\ry*\RH) -- (\TX+\TW,\Rz+\ry*\RH);
}

\fill[cpuBlue!55]   (\TX+\LW+1.5, \Rz+2*\RH+0.03) rectangle (\TX+\LW+3.50,\Rz+3*\RH-0.03);
\node[font=\large,black] at (\TX+\LW+2.5,\Rz+2.5*\RH) {$s$};

\fill[cpuBlue!55]   (\TX+\LW+3.55,\Rz+2*\RH+0.03) rectangle (\TX+\LW+6.25,\Rz+3*\RH-0.03);
\node[font=\large,black] at (\TX+\LW+5.05,\Rz+2.5*\RH) {$k_2$};

\fill[gpuGreen!80]  (\TX+\LW+1.60,\Rz+1*\RH+0.03) rectangle (\TX+\LW+3.50,\Rz+2*\RH-0.03);
\node[font=\large,black] at (\TX+\LW+2.55,\Rz+1.5*\RH) {$g_1$};

\fill[gpuGreen!80]  (\TX+\LW+4.55,\Rz+1*\RH+0.03) rectangle (\TX+\LW+5.55,\Rz+2*\RH-0.03);
\node[font=\large,black] at (\TX+\LW+5.05,\Rz+1.5*\RH) {$g_2$};

\fill[commPurple!80] (\TX+\LW,\Rz+0*\RH+0.03) rectangle (\TX+\LW+1.5,\Rz+1*\RH-0.03);
\node[font=\large,black] at (\TX+\LW+.75,\Rz+0.5*\RH) {$l_0$};

\draw[commOrange!60!black,line width=1.0pt]
  (\TX+\LW+0.02,\Rz+0*\RH+0.03) rectangle (\TX+\LW+1.50,\Rz+1*\RH-0.03);

\fill[commPurple!80] (\TX+\LW+3.50,\Rz+0*\RH+0.03) rectangle (\TX+\LW+4.50,\Rz+1*\RH-0.03);
\node[font=\large,black] at (\TX+\LW+4.0,\Rz+0.5*\RH) {$l_1$};

\fill[commOrange]    (\TX+\LW+6.25,\Rz+0*\RH+0.03) rectangle (\TX+\LW+7.0,\Rz+1*\RH-0.03);
\node[font=\large,black] at (\TX+\LW+6.625,\Rz+0.5*\RH) {$l_2$};
\draw[commOrange!60!black,line width=1.0pt]
  (\TX+\LW+6.25,\Rz+0*\RH+0.03) rectangle (\TX+\LW+7.0,\Rz+1*\RH-0.03);

\draw[causal]
  (\TX+\LW+6.9, \Rz+1*\RH-0.03)
  to[bend right=35]
  (\TX+\LW+6.25, \Rz+2.5*\RH-0.03);

\draw[causal]
  (\TX+\LW+5.05, \Rz+3*\RH-0.03)
  to[bend right=15]
  (\TX+\LW+2.50, \Rz+3*\RH-0.03);

\draw[causal]
  (\TX+\LW+1.50, \Rz+2.7*\RH-0.03)
  to[bend right=35]
  (\TX+\LW+0.75, \Rz+1*\RH-0.03);

\node[anchor=south, rotate=90, font=\small\bfseries, headerBg, inner sep=1.5pt]
  at (\TX+\LW-0.05, \Rz+1.5*\RH) {Rank 0};

\def\R1Y{-1.40}
\def\Ro{\R1Y}

\foreach \ry in {0,1,2,3}{
  \draw[gray!40,very thin] (\TX+\LW,\Ro+\ry*\RH) -- (\TX+\TW,\Ro+\ry*\RH);
}

\fill[cpuBlue!55]  (\TX+\LW+1.50,\Ro+2*\RH+0.03) rectangle (\TX+\LW+2.50,\Ro+3*\RH-0.03);
\node[font=\large,black] at (\TX+\LW+2.0,\Ro+2.5*\RH) {$s$};

\fill[cpuBlue!55]  (\TX+\LW+2.60,\Ro+2*\RH+0.03) rectangle (\TX+\LW+3.60,\Ro+3*\RH-0.03);
\node[font=\large,black] at (\TX+\LW+3.10,\Ro+2.5*\RH) {$k_2$};

\fill[gpuGreen!80] (\TX+\LW + 1.5,\Ro+1*\RH+0.03) rectangle (\TX+\LW+2.5,\Ro+2*\RH-0.03);
\node[font=\large,black] at (\TX+\LW+2.0,\Ro+1.5*\RH) {$g_1$};

\fill[gpuGreen!80] (\TX+\LW+4.50,\Ro+1*\RH+0.03) rectangle (\TX+\LW+5.20,\Ro+2*\RH-0.03);
\node[font=\large,black] at (\TX+\LW+4.85,\Ro+1.5*\RH) {$g_2$};

\fill[commPurple!80] (\TX+\LW,\Ro+0*\RH+0.03) rectangle (\TX+\LW+1.5,\Ro+1*\RH-0.03);
\node[font=\large,black] at (\TX+\LW+.75,\Ro+0.5*\RH) {$l_0$};

\draw[commOrange!60!black,line width=1.0pt]
  (\TX+\LW+0.03,\Ro+0*\RH+0.03) rectangle (\TX+\LW+1.50,\Ro+1*\RH-0.03);

\fill[commPurple!80] (\TX+\LW+2.50,\Ro+0*\RH+0.03) rectangle (\TX+\LW+4.50,\Ro+1*\RH-0.03);
\node[font=\large,black] at (\TX+\LW+3.50,\Ro+0.5*\RH) {$l_1$};

\fill[commOrange] (\TX+\LW+5.20,\Ro+0*\RH+0.03) rectangle (\TX+\LW+7.0,\Ro+1*\RH-0.03);
\node[font=\large,black] at (\TX+\LW+6.10,\Ro+0.5*\RH) {$l_2$};
\draw[commOrange!60!black,line width=1.0pt]
  (\TX+\LW+5.20,\Ro+0*\RH+0.03) rectangle (\TX+\LW+7.0,\Ro+1*\RH-0.03);

\draw[causal]
  (\TX+\LW+6.70, \Ro+1*\RH-0.03)
  to[bend right=35]
  (\TX+\LW+5.20, \Ro+1.5*\RH-0.03);

\draw[causal]
  (\TX+\LW+4.50, \Ro+1.5*\RH-0.03)
  to[bend right=35]
  (\TX+\LW+3.55, \Ro+1*\RH-0.03);

\draw[causal]
  (\TX+\LW+3.50, \Ro+1*\RH-0.03)
  to[bend right=35]
  (\TX+\LW+2.50, \Ro+1.5*\RH-0.03);

\draw[causal]
  (\TX+\LW+1.53, \Ro+1.5*\RH-0.03)
  to[bend right=35]
  (\TX+\LW+.50, \Ro+1*\RH-0.03);

\node[anchor=south, rotate=90, font=\small\bfseries, headerBg, inner sep=1.5pt]
  at (\TX+\LW-0.05, \Ro+1.5*\RH) {Rank 1};

\def\RankSep{3.6}
\def\SC{1.0}
\def\VS{0.58}
\def\DX{-0.7}
\def\DY{-4.15}

\draw[gray!50, dashed, line width=0.6pt]
  (\DX+\RankSep/2, \DY-0.05) -- (\DX+\RankSep/2, \DY+4.0*\VS+0.06);

\node[font=\normalsize\bfseries, cpuBlue]
  at (\DX, \DY+4.2*\VS) {Rank 0};

\node[commOC] (r0C2)  at (\DX,      \DY+3.5*\VS) {$l_2$};
\node[cpuC]   (r0KL)  at (\DX+\SC,  \DY+2.75*\VS) {$k_2$};
\node[gpuC]   (r0G2)  at (\DX-\SC,  \DY+2.75*\VS) {$g_2$};
\node[commC]  (r0C1)  at (\DX,      \DY+2.0*\VS) {$l_1$};
\node[cpuC]   (r0SS)  at (\DX+\SC,  \DY+1.25*\VS) {$s$};
\node[gpuC]   (r0G1)  at (\DX-\SC,  \DY+1.25*\VS) {$g_1$};
\node[commC]  (r0C0m) at (\DX,      \DY+0.5*\VS) {$l_0$};

\draw[causal] (r0C2)  -- (r0KL);
\draw[dagedge] (r0C2)  -- (r0G2);
\draw[dagedge] (r0G1)  -- (r0C0m);

\draw[dagedge] (r0G2)  -- (r0C1);
\draw[dagedge] (r0C1)  -- (r0G1);

\draw[dagedge] (r0C2)  -- (r0C1);
\draw[causal] (r0KL)  -- (r0SS);
\draw[causal] (r0SS)  -- (r0C0m);
\draw[dagedge] (r0G2)  -- (r0G1);
\draw[dagedge] (r0C1)  -- (r0C0m);

\def\RX{\DX+\RankSep}

\node[font=\normalsize\bfseries, cpuBlue]
  at (\RX, \DY+4.2*\VS) {Rank 1};

\node[commOC] (r1C2)  at (\RX,      \DY+3.5*\VS) {$l_2$};
\node[cpuC]   (r1KL)  at (\RX+\SC,  \DY+2.75*\VS) {$k_2$};
\node[gpuC]   (r1G2)  at (\RX-\SC,  \DY+2.75*\VS) {$g_2$};
\node[commC]  (r1C1)  at (\RX,      \DY+2.0*\VS) {$l_1$};
\node[cpuC]   (r1SS)  at (\RX+\SC,  \DY+1.25*\VS) {\textit{s}};
\node[gpuC]   (r1G1)  at (\RX-\SC,  \DY+1.25*\VS) {$g_1$};
\node[commC]  (r1C0m) at (\RX,      \DY+0.5*\VS) {$l_0$};

\draw[dagedge] (r1C2)  -- (r1KL);
\draw[causal] (r1C2)  -- (r1G2);
\draw[causal] (r1G2)  -- (r1C1);
\draw[dagedge] (r1C2)  -- (r1C1);
\draw[dagedge] (r1KL)  -- (r1SS);
\draw[dagedge] (r1G2)  -- (r1G1);
\draw[dagedge] (r1C1)  -- (r1C0m);

\draw[dagedge] (r1SS)  -- (r1C0m);

\draw[causal] (r1C1)  -- (r1G1);
\draw[causal] (r1G1)  -- (r1C0m);

\def\SWL{0.36}    %
\def\SWH{0.13}    %
\def\LX{5.10}     %
\def\LRS{0.78}    %
\def\LGY{0.30}    %

\fill[cpuBlue!55]    (\LX,\LGY-\SWH) rectangle (\LX+\SWL,\LGY+\SWH);
\node[anchor=west,font=\small] at (\LX+\SWL+0.10,\LGY) {CPU event};

\fill[gpuGreen!80]   (\LX,\LGY-\LRS-\SWH) rectangle (\LX+\SWL,\LGY-\LRS+\SWH);
\node[anchor=west,font=\small] at (\LX+\SWL+0.10,\LGY-\LRS) {GPU Comp.};

\fill[commPurple!80] (\LX,\LGY-2*\LRS-\SWH) rectangle (\LX+\SWL,\LGY-2*\LRS+\SWH);
\node[anchor=west,font=\small] at (\LX+\SWL+0.10,\LGY-2*\LRS) {GPU Comm.};

\fill[commOrange]    (\LX,\LGY-3*\LRS-\SWH) rectangle (\LX+\SWL,\LGY-3*\LRS+\SWH);
\node[anchor=west,font=\small] at (\LX+\SWL+0.10,\LGY-3*\LRS) {Start Comm.};

\draw[causal] (\LX,\LGY-4*\LRS) -- (\LX+\SWL+0.14,\LGY-4*\LRS);
\node[anchor=west,font=\small] at (\LX+\SWL+0.24,\LGY-4*\LRS) {Critical Path};

\end{tikzpicture}}
\caption{Critical paths of collective $l_2$ on two ranks. In the traces (top), each rank has events on a CPU thread, a compute stream ($g_1$, $g_2$), and a communication stream ($l_0$, $l_1$, $l_2$). On the CPU, $s$ is a \texttt{cudaStreamSynchronize} call and $k_2$ launches $l_2$. Red arrows follow the critical path of each rank backward from $l_2$. Rank~0 enters $l_2$ last, as its shorter $l_2$ kernel shows, because its CPU launches $l_2$ late. Its path therefore runs through $k_2$ and $s$ on the CPU, whereas the path of rank~1 stays on the GPU. In the dependency graphs (bottom), the walk of rank~1 stops at the previous collective $l_1$, but the walk of rank~0 bypasses $l_1$ and stops at $l_0$. We extend the path of rank~1 to $l_0$, the most recent collective that both paths contain.}
\label{fig:algo}
\vspace{-1mm}
\end{figure}

\subsection{Critical-Path Extraction}
\label{sec:critical-path}
\myparab{Traversal.} Consider a collective $l$ issued by a parallelism group (\eg TP). On every rank, we start at the node of $l$ in the dependency graph and walk backward. We move to the parent node that finished last, because the current node could not start until all of its parents finished. If the parent on the same stream or thread as the current node finished within a short time (\eg $\leq 3~\mu$s) of the last-finishing parent, we move to the same-stream parent instead. In-order execution on streams makes the same-stream edge a definite dependency, whereas edges from other streams are inferred. This walk stops when it reaches another collective on the stream of $l$, which we call the \emph{ancestor} of $l$ on that rank. The nodes traversed between the ancestor and $l$ form the critical path of the rank.

\begin{figure*}[t]
\centering
\includegraphics[width=\linewidth]{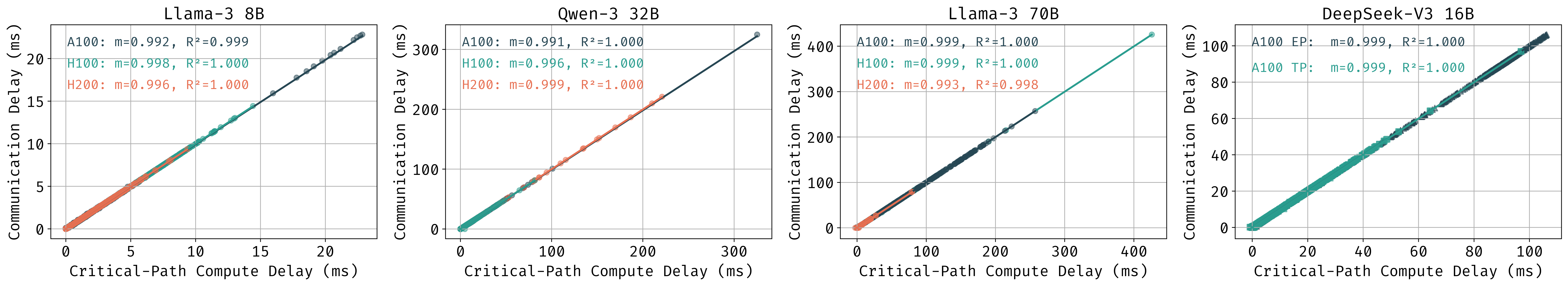}
\caption{Measured synchronization delay of the first rank to enter, $\tau_{\text{first}}$ (y-axis), vs.\ the compute critical-path difference $P_{\text{last}} - P_{\text{first}}$ inferred by our algorithm (x-axis). Each point is one collective.}
\label{fig:synctax-scatter}
\end{figure*}

\myparab{Alignment.} Comparing critical paths across ranks requires a common reference point across ranks since timestamps from different ranks' traces are not comparable. The end of a collective provides this reference, since all ranks finish a collective at nearly the same time. However, the traversals of different ranks can stop at different ancestors, \eg when the path of one rank passes through a CPU event (Figure~\ref{fig:algo}). To address this, we extend each path backward, one collective at a time, until all ranks reach the most recent collective that every path contains, \ie their \emph{lowest common ancestor}.

\myparab{Attribution.} We compare the extended paths of the last and first ranks to enter $l$. We match each event on the path of the last rank to the event with the same name and position on the path of the first rank and take the difference in their durations. If the path of the first rank lacks a matching event, the difference is the full duration of the event on the last rank. The event with the largest difference is the primary source of the delay at $l$. Unlike profilers that report the longest events on a rank, this comparison ignores events with the same duration on every rank (even if it is long) while focusing on events with high \emph{variation} across ranks.

\subsection{Evaluation and Results}
\label{sec:attribution-results}

If the critical paths extracted by our algorithm are correct then we expect that the difference in durations of critical paths will predict the measured synchronization delay in our experiments (\S\ref{sec:characterization}). Our evaluation demonstrates that our algorithm attributes nearly all of the synchronization delay to events on the critical path.

\myparab{Critical paths account for the delay.} Let the \emph{path duration} $P$ of a rank be the total duration of the events on its aligned critical path, starting from the lowest common ancestor. A rank enters $l$ at that common instant plus $P$, so $P_{\text{last}} - P_{\text{first}}$ equals the difference between the entry times of the last and first ranks to enter. In \S\ref{sec:profiling} we measured this difference independently from NCCL kernel durations as the synchronization delay of the first rank, $\tau_{\text{first}} = d_{\text{first}} - d_{\text{last}}$. If the recovered critical paths are correct, then
\begin{equation}
\tau_{\text{first}} \;\approx\; P_{\text{last}} - P_{\text{first}}.
\label{eq:scatter}
\end{equation}

We run our algorithm on all collectives of every workload in Table~\ref{tab:workloads}. Figure~\ref{fig:synctax-scatter} plots the two sides of Eq.~\ref{eq:scatter}. We find that the best-fit line for every workload has a slope of ${\approx}1$ and a coefficient of determination ($R^{2}$) of ${\approx}1$, \ie path differences explain nearly all of the variation in the measured delay. This agreement implies that ranks finish each collective together, so data transfer does not contribute to the delay, and that the critical paths account for all the time before each collective. For fewer than 5\% of collectives, the paths of the first and last ranks differ in type (CPU vs.\ GPU), and comparing them requires adjusting for gaps between GPU kernels that the profiler does not record. We evaluate these collectives separately and find similar accuracy (Appendix~\ref{sec:algo}).

\myparab{Both the traversal rule and the inferred edges are necessary.} To understand each design choice in our algorithm, we rerun our algorithm on the Llama-3 70B traces from H200 GPUs with one design choice changed at a time (Appendix~\ref{sec:neg-controls}). A traversal that moves to the longest-running parent, instead of the parent that finished last, yields $R^{2}{=}0.204$, and a dependency graph without inferred cross-stream edges achieves $R^{2}{=}0.646$, compared with $R^{2}{=}0.998$ for our algorithm. Both variants select parents that finish far earlier, so their paths miss the time spent waiting on the true dependency.

\begin{figure}[h]
\centering
\includegraphics[width=\linewidth]{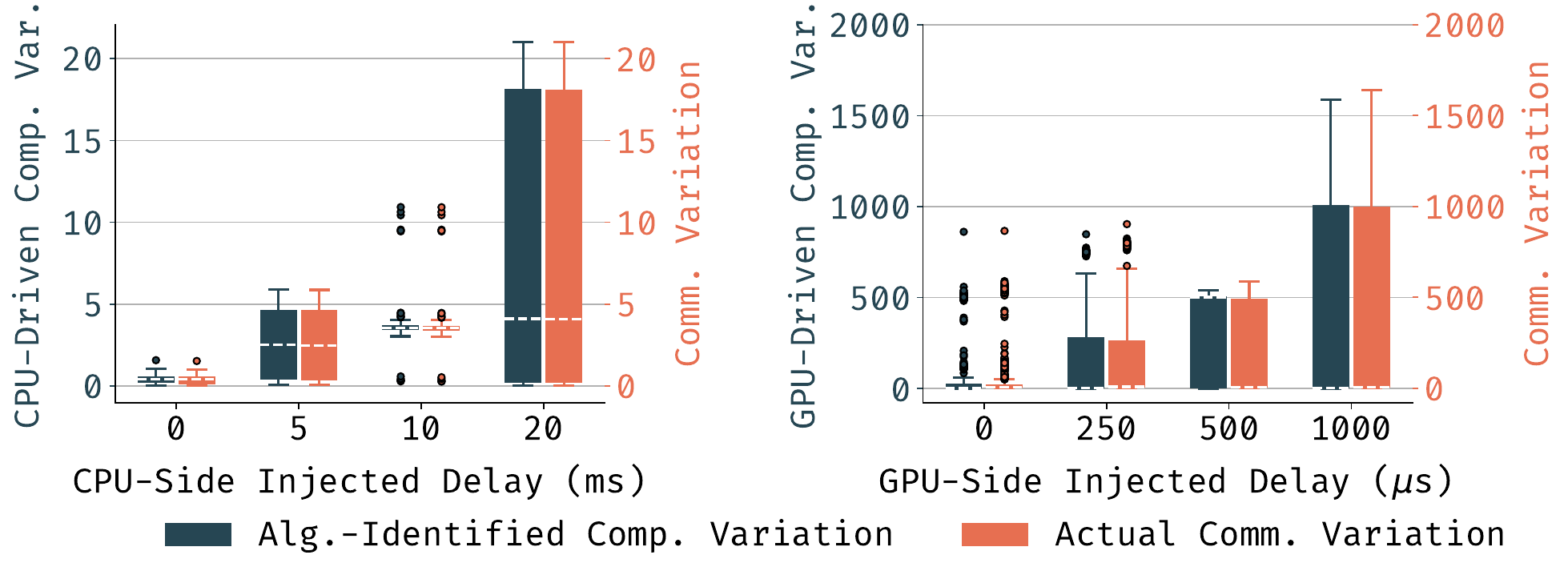}
\caption{Delay attributed by our algorithm vs.\ measured synchronization delay, with delays injected on CPU (left) and GPU (right).}
\label{fig:delay-injection}
\end{figure}

\myparab{Interventional validation.} The agreement with Eq.~\ref{eq:scatter} provides observational evidence in Figure~\ref{fig:synctax-scatter}. We also evaluate our algorithm by injecting delays of known size into known events and checking that our algorithm attributes the resulting synchronization delay to them. We train Llama-3 8B with TP on four A100s. On the GPU, we insert a kernel that busy-waits for 0--1000~$\mu$s through a custom \texttt{TorchDispatchMode} handler. Our algorithm identifies the injected kernel on the critical path, and the delay it attributes matches the measured synchronization delay as the injection grows (Figure~\ref{fig:delay-injection}, right). On the CPU, we delay the launches of GEMM kernels by 0--20~ms. Launches normally run well ahead of the GPU, so only millisecond-scale delays move the critical path from the GPU to the CPU. As the injected delay grows, the delay attributed to CPU events grows and matches the measured delay (Figure~\ref{fig:delay-injection}, left).

\myparab{GEMMs take the blame.} We classify the events on each critical path by kernel name (\eg \texttt{gemm} and \texttt{nvjet} denote GEMMs). For 93.7\% of collectives, the critical path of the last rank to enter lies entirely on the GPU, \ie CPU-side events rarely cause the synchronization delay. Table~\ref{tab:top-kernels} breaks down the GPU-side difference between the paths of the first and last ranks by kernel class. GEMM kernels account for 78.0\% of this difference.

\begin{table}[h]
\centering
\resizebox{\columnwidth}{!}{
\begin{tabular}{lcccccc}
\toprule
& \textbf{GEMM} & FlashAttention & FSDP & Norm & Concat/Split \\
\midrule
\% of GPU-side difference & \textbf{78.0\%} & 15.4\% & 4.3\% & 1.2\% & 0.6\% \\
\bottomrule
\end{tabular}
}
\caption{\small Kernel classes that account for the GPU-side difference between the critical paths of the first and last ranks to enter.}
\label{tab:top-kernels}
\end{table}

\begin{figure*}[t]
\centering
\begin{subfigure}[b]{0.565\textwidth}
\includegraphics[width=\linewidth]{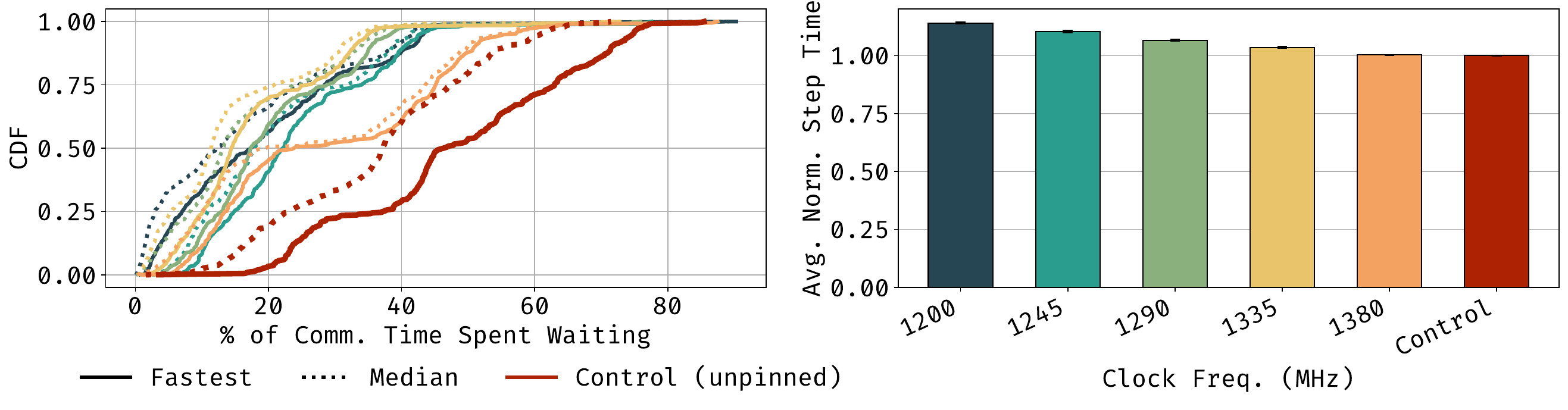}
\caption{Clock pinning}
\label{fig:clocks}
\end{subfigure}%
\hfill
\begin{subfigure}[b]{0.415\textwidth}
\includegraphics[width=\linewidth]{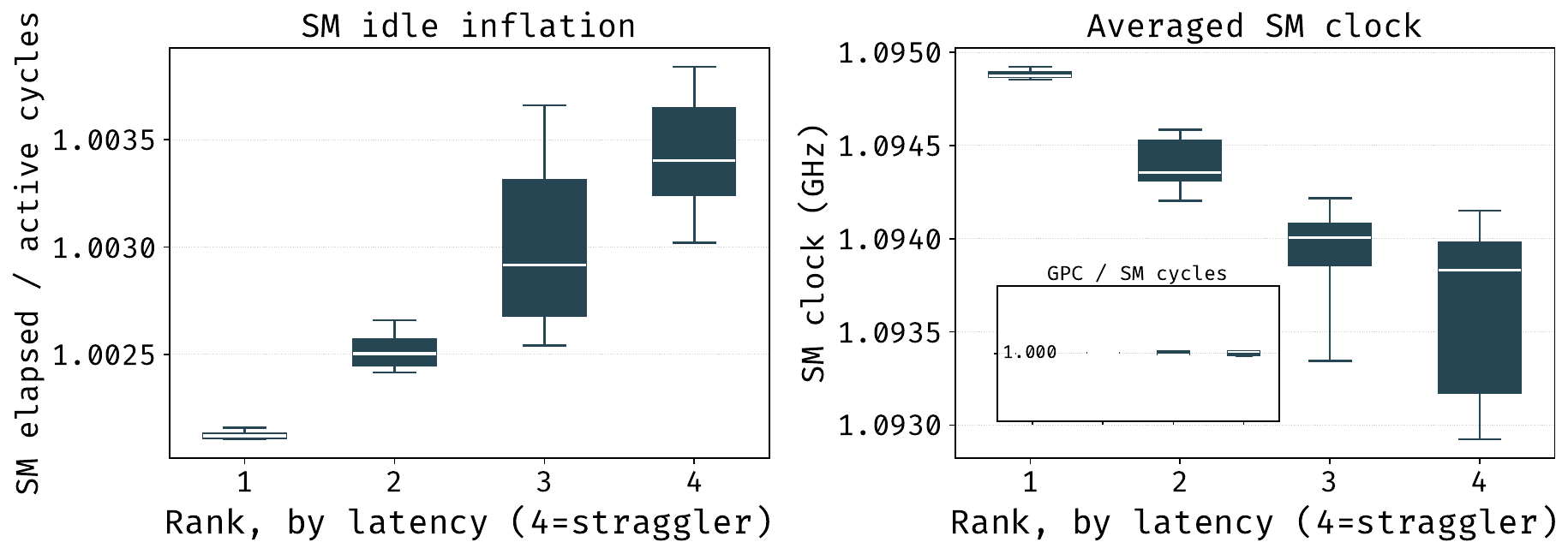}
\caption{Hardware counters}
\label{fig:ncu}
\end{subfigure}
\caption{Sources of GEMM variation. (a) Synchronization delay (left) and step time (right) across pinned SM clock frequencies, normalized to an unpinned run. (b) \texttt{ncu} counters for the \texttt{ampere\_\allowbreak sgemm\_\allowbreak 128x128\_\allowbreak nn} kernel, with ranks ordered from fastest (1) to slowest (4).}
\label{fig:gemm-sources}
\end{figure*}
\subsection{Sources of GEMM-Kernel Variation}
\label{sec:gemm-variation}

We attribute most of the synchronization delay to GEMM kernels whose runtimes differ across ranks (\S\ref{sec:attribution-results}). This variation is unexpected since each rank executes GEMMs of identical dimensions on dense matrices, performing identical work. We investigate the reasons for variation in the runtime of GEMM kernels across ranks.

\myparab{Clock frequency.} Each GPU sets its clock frequency independently through dynamic voltage and frequency scaling (DVFS)~\cite{bharadwaj2023predict}. Due to this, the same GEMM can run at different speeds on different ranks. We pin the SM clocks of four A100 GPUs in a server to the same frequency (\texttt{nvidia-smi~-lgc}) and train Llama-3 8B with TP for 30 iterations. We find that while pinning reduces the synchronization delay, it does not eliminate it entirely (Figure~\ref{fig:clocks}, left). Lower clock frequencies reduce the delay the most but slow down training by up to 14\% (Figure~\ref{fig:clocks}, right). Thus, clock variation only explains part of the synchronization delay.

\myparab{Hardware counters.} We collect hardware counters for one GEMM kernel (\texttt{ampere\_\allowbreak sgemm\_\allowbreak 128x128\_\allowbreak nn}) with NVIDIA's \texttt{ncu} profiler~\cite{ncu} over 70 iterations of Llama-3 8B on a node with 4 A100 GPUs. We order the ranks from fastest to slowest for each kernel instance. Active SM cycles agree across ranks to within $10^{-5}$, confirming identical work per rank. However, slower ranks spend more cycles with no warp resident on the SM (Figure~\ref{fig:ncu}, left), and the slowest rank averages an SM clock 1.3~MHz lower than the fastest (Figure~\ref{fig:ncu}, right), even with \texttt{ncu}'s default clock control. The slowest rank performs the same work but loses time to idle cycles and brief drops in frequency. Other hardware counters during GEMM profiling did not reveal any correlation with the synchronization delay. Thus, we conclude that the variation in GEMM runtimes is difficult to control and likely intrinsic to distributed execution.

\section{Modeling the Synchronization Delay}
\label{sec:tax-model}
In \S\ref{sec:attribution}, we found that cross-rank variation in GEMM runtimes accounts for most of the synchronization delay, even though every rank performs identical work. Pinning GPU clock frequencies reduces the delay but also reduces performance while still failing to eliminate the delay. Moreover, the slowest rank changes from one collective to the next (\S\ref{sec:characterization}), so the delay does not stem from a single device that can be replaced. Therefore, synchronization delay seems unavoidable even in homogeneous scale-up domains. Now, we develop an analytical model of the synchronization delay from cross-rank variation in GEMM runtimes and use it to predict the delay in larger domains to inform the design of future scale-up domains. Since other sources of variation (\eg EP token imbalance) can only add to the delay, our model provides a lower bound.

\begin{figure*}[t]
\centering
\includegraphics[width=0.7\linewidth]{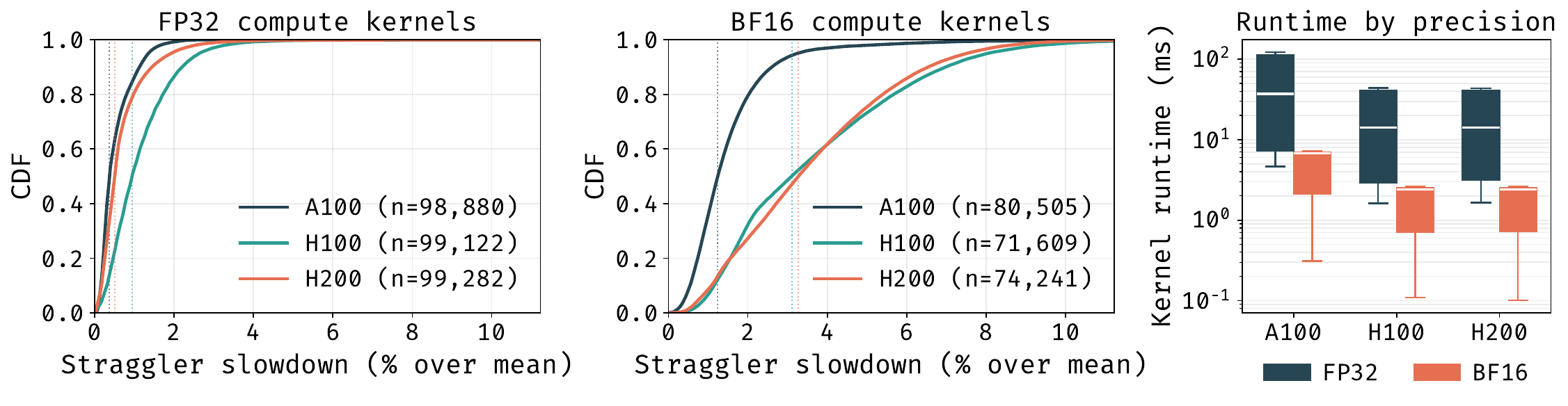}
\vspace{-2mm}
\caption{Variation across ranks in all $n$ instances of each GPU compute kernel (attention, GEMM) across workloads. Straggler slowdown reflects the maximum value's percent increase in runtime over the mean for each instance of the kernel (\ie $(M_{c,j} - \bar{X}_{c,j})/\bar{X}_{c,j}$).}
\label{fig:comp-skews}
\end{figure*}

\subsection{Applying Extreme Value Theory}
\label{sec:evt}
The synchronization delay is set by the slowest of $n$ ranks to arrive at each collective. Our attribution algorithm allows us to be more precise: the delay is set by the largest GEMM execution time across ranks before each collective. We study this using extreme value theory (EVT), a statistical framework for characterizing maxima.

\myparab{Why EVT?} EVT is widely used across disciplines to model the maximum of many random samples and how it grows with the number of samples~\cite{coles2001introduction, katz2002statistics}. Alternatively, we could fit a distribution $F$ to the measured GEMM runtimes and compute the distribution of the maximum across $n$ ranks as $F^{n}$. However, $F^{n}$ for large $n$ depends almost entirely on the upper tail of $F$, which the fit captures least accurately, so small fitting errors grow with $n$. In contrast, EVT provides a powerful and general result that under broad conditions, the distribution of normalized sample maxima, \emph{regardless of underlying parent distribution}, converges to the generalized extreme value distribution (GEVD)~\cite{coles2001introduction}. We model the GEMM critical-path computation time across $n$ ranks as samples $X_1,\ldots,X_n$, so that the synchronization delay is governed by their maximum $M_n{=}\max_i X_i$.

We fit the GEVD to empirical maxima, verify its conditions, and validate that it accurately predicts the synchronization delay (\S\ref{sec:model-eval}). We extrapolate to larger $n$ using the GEVD's exact closure property under repeated maximization~\cite{coles2001introduction} (Appendix~\ref{sec:frechet}). We use EVT to provide a \textbf{conservative estimate} of the delay, \ie excluding less predictable phenomena (\eg MoE load imbalance) that occur on top of routine compute variation. The GEVD family comprises three shapes: Gumbel, Fréchet, Weibull, \ie light-tailed, heavy-tailed, and bounded distributions, respectively. While a Weibull model requires a finite upper bound, GPU execution provides no such strict bound, making it less appropriate for our setting. Thus, we focus on the light-tailed Gumbel because its logarithmic scaling provides a conservative estimate compared to the polynomial scaling of the heavy-tailed Fréchet. Our empirical data is better fit by the Fr\'{e}chet distribution (Appendix~\ref{sec:frechet}), so our results provide a \emph{lower} bound on the delay at scale.

\myparab{Max-stability.} A key property of the GEVD is \emph{max-stability}: the maximum of $k$ i.i.d. samples from a GEVD is itself GEVD with the same shape, allowing us to extrapolate to larger domains~\cite{coles2001introduction} (Appendix~\ref{sec:frechet}). (Samples are i.i.d. as they are the same kernels on different GPUs, as shown in \S\ref{sec:fitting}.) If $X{\sim}\text{Gumbel}(\mu, \sigma)$, then $\mathbb{E}[X]{=}\mu{+}\gamma\sigma$, where $\gamma{\approx}0.577$ is the Euler-Mascheroni constant~\cite{coles2001introduction}. By max-stability~\cite{coles2001introduction}, if we fit on $m$ ranks (\S\ref{sec:fitting}), the expected maximum in a domain of $n{=}km$ GPUs, \ie $k$ such samples, is
\begin{equation}
\mathbb{E}[M_n] = \mathbb{E}\!\left[\max_{u=1}^k \Big(\max_{i=1}^m X_i\Big)\right]
= \mu + \sigma (\ln(k) + \gamma).
\label{eq:gumbel-ev}
\end{equation}
Max-stability lets us predict the expected maximum GEMM runtime across ranks, and from it the synchronization delay, at larger scales without more measurements. Eq.~\ref{eq:gumbel-ev} shows that \emph{at best} (\ie Gumbel scaling), the delay grows \emph{logarithmically} with domain size.

\subsection{Fitting the Empirical Data}
\label{sec:fitting}
Across our workloads, we fit the GEVD on all compute blocks, \ie distinct sequences of GEMM kernels between consecutive TP collectives (Appendix~\ref{sec:comp-blocks}). We fit on compute blocks instead of individual GEMM kernel runtimes because each compute block can consist of several back-to-back GEMM kernels, and modeling skew at the kernel level would require capturing dependence across them. For example, assuming independence may \emph{underestimate} compute skew, as kernels can consistently run longer on one rank during a compute block. While we focus here on TP collectives to establish a \emph{lower bound}, \ie fundamental variation despite identical work, our model only requires per-rank GEMM times between consecutive collectives and thus applies to other parallelisms (as validated in \S\ref{sec:model-eval}). Parallelisms with unequal work, \ie EP, only increase this variation.

We first normalize GEMM compute-block execution times into $z$-scores (\ie standard scores) that eliminate workload and architecture-specific effects. The GEVD is fit on the maxima of samples, each of a fixed size $m$. In our case, a sample is the set of execution times across ranks participating in the same collectives (\ie same TP group) for each compute-block in the trace. Thus, we fix $m{=}4$, since this is the smallest TP group size across all our workloads. (For Qwen-3 32B with 8-way TP, we assign odd ranks to one ``sample'' and even ranks to the other to ensure $m{=}4$.) Then, for each distinct compute block $c$, we iterate over all its instances $j$ and calculate the $z$-score of the maximum value among the $m$ ranks as $z_{c,j} = \tfrac{M_{c,j} - \bar{X}_{c,j}}{s_{c,\text{pooled}}}$,
where $M_{c,j}$ is the maximum value (runtime) of the $j$-th instance of compute block $c$ across the $m$ ranks, $\bar{X}_{c,j}$ is the mean, and $s_{c,\text{pooled}}$ is the pooled standard deviation across all instances of $c$ (Appendix~\ref{sec:pooled-std}).

Then, we fit the GEVD to these $z$-scores using maximum likelihood estimation (MLE), the standard for model fitting, and provide the fits in Appendix~\ref{sec:frechet}. We confirm key EVT conditions and validate its scaling within our empirical data below.

\begin{figure}[h]
\centering
\begin{subfigure}[b]{0.49\linewidth}
  \centering
  \includegraphics[width=\linewidth]{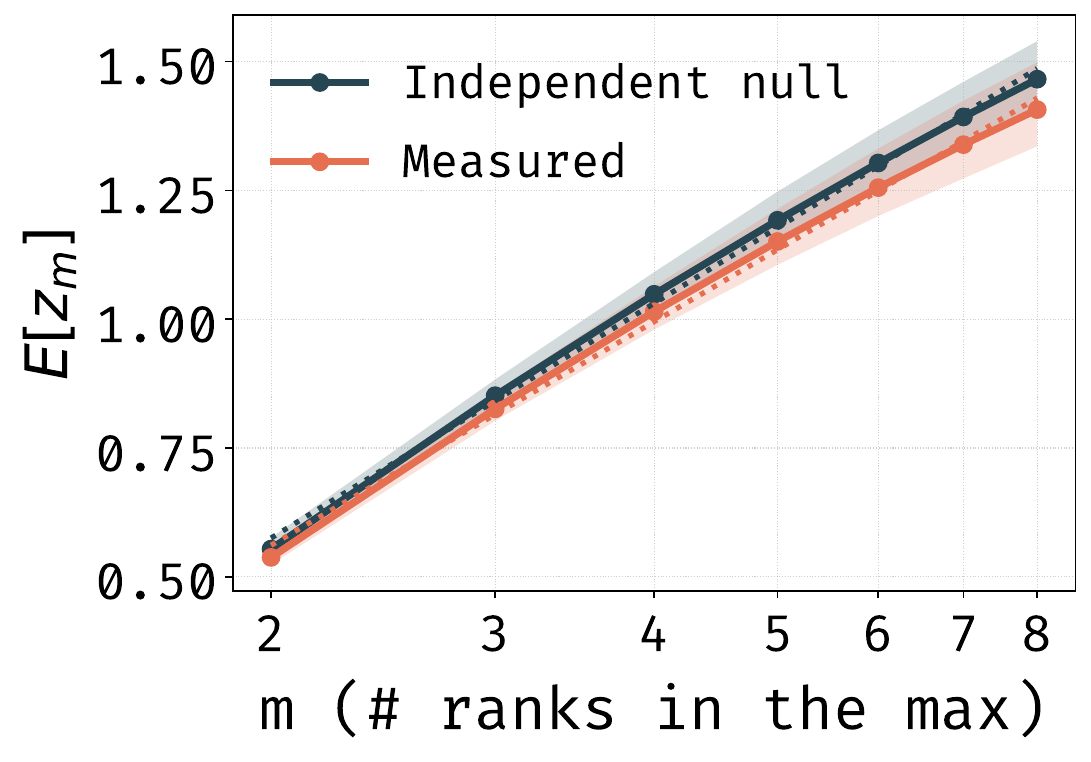}
  \vspace{-2em}
  \caption{Independent compute times.}
  \label{subfig:main-scaling}
\end{subfigure}
\hfill
\begin{subfigure}[b]{0.49\linewidth}
  \centering
  \includegraphics[width=\linewidth]{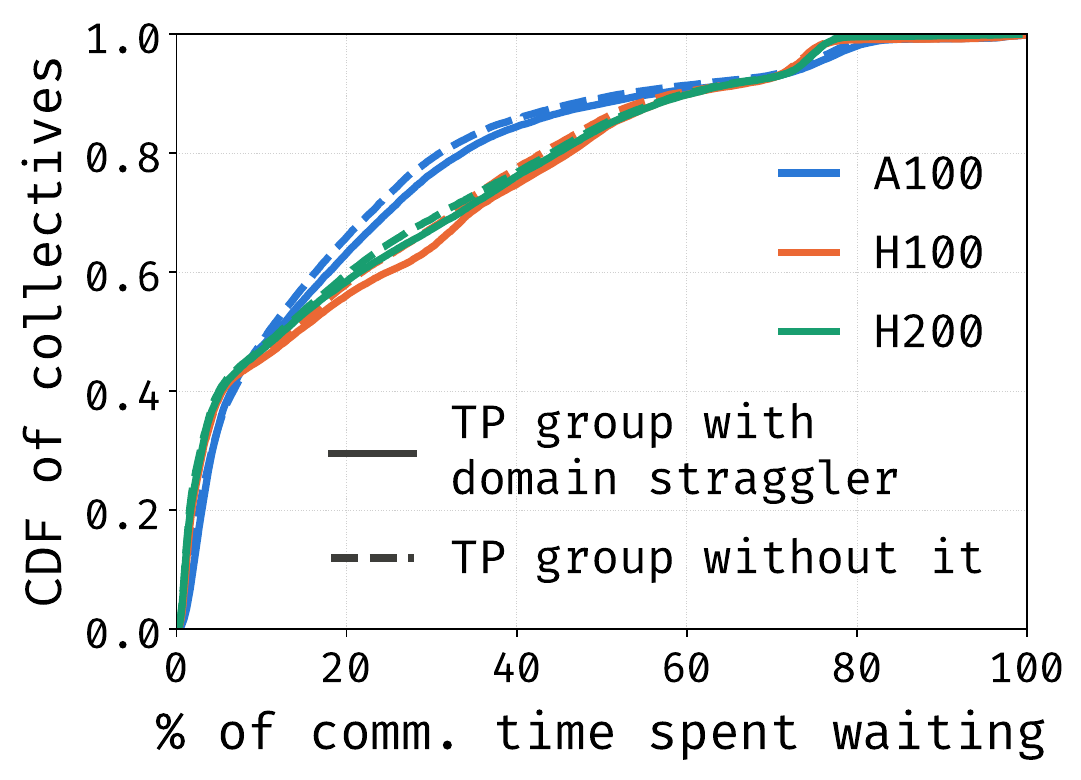}
  \vspace{-2em}
  \caption{Domain-wide delay.}
  \label{subfig:main-domain}
\end{subfigure}
\caption{(a) Expected maximum $z$-score across subsets of $m$ ranks, measured vs. an independent baseline. Shaded regions show the central 95\% of resamples. (b) Synchronization delay in Llama-3 70B TP groups with vs. without the domain's slowest GPU.}
\label{fig:indep-coupling}
\end{figure}

\myparab{Independence.} We compute the expected maximum $z$-score over every subset of $m{\in}[2,8]$ ranks for each compute block over all our 8-GPU workloads (Qwen-3 32B, Llama-3 70B). We then construct an independent baseline that randomly permutes each rank's executions of the same compute block, \ie without cross-rank temporal correlation. Figure~\ref{fig:indep-coupling}a shows that the measured maximum $z$-score grows similarly to the independent baseline with the block size $m$ (ratio of their log-linear slopes in $m$ is $\lambda{=}0.95$), satisfying EVT's independence conditions. To quantify dependence on which compute blocks we profiled, Figure~\ref{fig:indep-coupling}a also shows a 95\% confidence interval, obtained by resampling compute blocks with replacement 500 times and recomputing each curve.

To extrapolate, we fix the GEVD's shape parameter as $\xi{=}0$ for the Gumbel distribution, for the \emph{most conservative estimate} of the synchronization delay. (Gumbel always scales more slowly than Fr\'{e}chet~\cite{coles2001introduction}.) While our model predicts a $z$-score for the expected maximum runtime of a compute block $c$ across ranks, we convert it back into time units by inverting standardization. Averaging Eq.~\ref{eq:synctax} over ranks, the delay is the expected gap between the slowest and average rank's compute-block execution time. Given a compute block $c$ with mean runtime $a{\approx}\bar{X}_c$ and standard deviation $b{\approx}s_{c,\text{pooled}}$, we calculate the synchronization delay for an $n$-GPU domain as
\begin{equation}
    \tau = \mathbb{E}[M_n] - \mathbb{E}[\bar{X}] = b(\mu + \sigma(\gamma + \ln(k))).
\label{eq:sync-tax}
\end{equation}
$\mu$ and $\sigma$ are the fitted Gumbel parameters, $b$ is an estimate of the specific kernel or compute block's standard deviation ($s_{c,\text{pooled}}$), and $k{=}n{/}m$ where $m$ is the block size used for fitting. Figure~\ref{fig:comp-skews} shows the distribution of the compute-block maxima ($M_{c,j}$) used to fit $z$-scores. BF16 kernels on newer GPUs (H100/H200) have heavier tails while FP32 kernels have lighter tails but longer runtimes.

\myparab{Domain-wide delay.} With hybrid parallelism, each parallelism group may run on a subset of the $n$ GPUs in the domain. However, collectives in another parallelism dimension propagate the delay, \eg a straggler in one TP group delays its FSDP peers, which then delay their own TP groups. As a result, each group often suffers from a delay set by the slowest GPU in the domain. We observe this empirically in the Llama-3 70B traces, as Figure~\ref{fig:indep-coupling}b shows that the TP group without the domain's slowest GPU still experiences 92--97\% of the median delay of the group with it.

\subsection{Evaluating the Fitted Model}
\label{sec:model-eval}
We evaluate how well our analytical model predicts the empirical synchronization delay. For the largest workloads (Qwen-3 32B, Llama-3 70B) we compare our model's prediction of the delay to the mean empirical synchronization delay at the collective. Our evaluation captures (1) if variation in GEMM compute times across ranks explains the synchronization delay, \emph{and} (2) if our model accurately predicts the synchronization delay by analytically modeling GEMM variation with EVT. Keeping with our conservative modeling, we constrain our evaluation to those collective operations for which our algorithm (\S\ref{sec:attribution}) traversed GPU-only paths with GEMMs, thereby excluding rare but extreme CPU-driven sources of variation.

\begin{figure}[h]
\centering
\includegraphics[width=\linewidth]{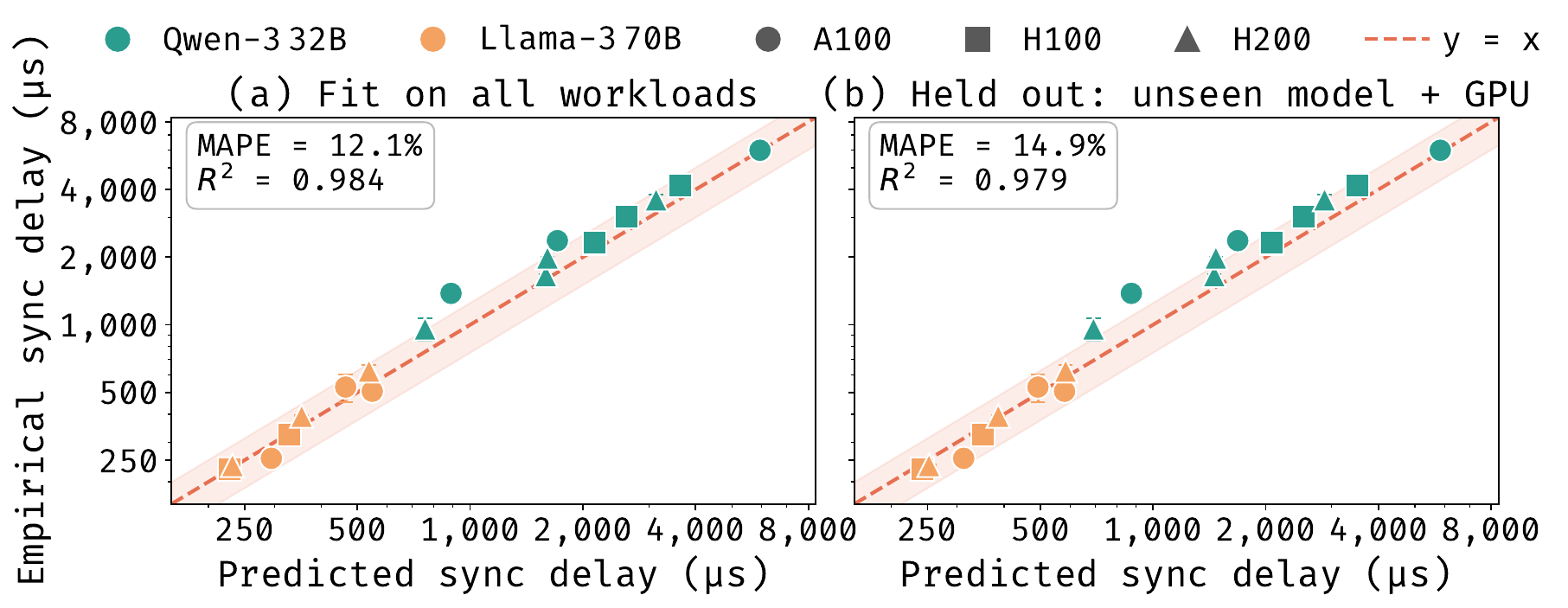}
\caption{Comparison of our (Gumbel) model's prediction to the empirical mean synchronization delay across all collectives following each unique compute block (data point). (Mappings from compute block ID to sequence of GEMMs are in Appendix~\ref{sec:comp-blocks}.) Left fits one global distribution whose parameters are later used in \S\ref{sec:bw-model} while right performs leave-one-out validation. Error bars are standard error of the mean, and shaded area is $\pm 25\%$, with only some Qwen workloads falling outside due to expected conservatism of the Gumbel.}
\label{fig:gumbel-val}
\end{figure}

Figure~\ref{fig:gumbel-val} (left) shows that our analytical model, based solely on GEMM variation, can accurately predict the mean synchronization delay across workloads and GPU models. While we fit the Gumbel using $m{=}4$ block size, the Qwen workloads use a TP size of 8 (so we plug in $k{=}8{/}4{=}2$), confirming that our model predicts the delay at scales larger than our fitted data. We also perform leave-one-out cross validation by fitting the Gumbel distribution on all workloads besides the one being evaluated in Figure~\ref{fig:gumbel-val} (right). The high $R^{2}$ and low mean absolute percentage error (MAPE) indicate our analytical model's accuracy and generalizability.

\begin{figure}[h]
\centering
\includegraphics[width=0.9\linewidth]{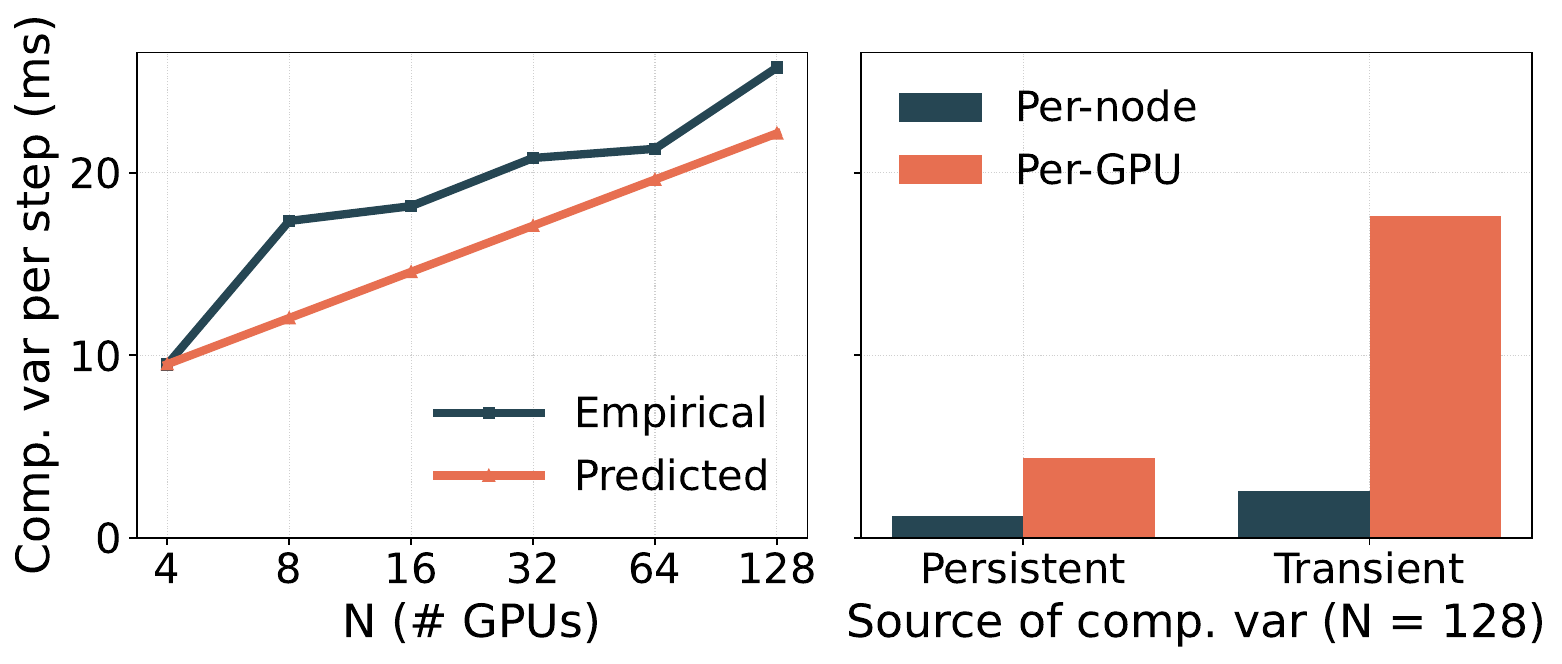}
\caption{Variation in compute-block runtimes across GPUs. Left compares our analytical model's prediction, with parameters fitted from the 4-GPU run, to the measured variation as the job size grows. Right shows that variation in the largest job ($N{=}128$ GPUs) is primarily driven by \emph{transient} variation at GPU (not node) granularity.}
\label{fig:scale-val}
\end{figure}

\myparab{Large-scale evaluation.} To assess whether our analytical model can be used to draw conclusions about the delay at larger scales, we run the Llama-3 8B workload across multiple nodes (each with 4 40-GB A100 GPUs) of the Perlmutter supercomputer, ranging from 4 GPUs (1 node) to 128 GPUs (32 nodes). In each job, we use TP within the node and FSDP between nodes. We fit the analytical model on \emph{only} the 4-GPU run and then use it to predict the compute variation at larger scales. Figure~\ref{fig:scale-val} (left) shows that our analytical model's predicted variation, simply using the mean and variance of GEMM kernels as measured on a 4-GPU run and applying EVT, can effectively capture compute variation at 128-GPU scale. The measured trend confirms that logarithmic scaling of compute variation (via the Gumbel model) provides a conservative and accurate framework to quantify the delay in large-scale GPU workloads. Figure~\ref{fig:scale-val} (right) confirms that this compute variation is driven by \emph{transient} GPU variation, rather than persistent offsets specific to a particular node or GPU or transient variation at node granularity.

\section{Implications for Scale-Up Design}
\label{sec:bw-model}

In this section, we show how accounting for the synchronization delay shifts the design of future scale-up domains. We incorporate our analytical model of the tax (\S\ref{sec:tax-model}) into ASTRA-sim~\cite{won2023astra}, a popular multi-GPU simulator, and explore how it changes both the optimal scale-up domain size and per-GPU interconnect bandwidth.

\subsection{Satisfying the communication budget}
\label{subsec:budget}
Understanding how synchronization impacts scale-up design decisions first requires defining an objective that architects should optimize for. Like prior ML co-design work, we use end-to-end training step time~\cite{won2024libra,isaev2023calculon,hsia2024mad}. We model the step time in a scale-up domain of $n$ GPUs with per-GPU scale-up bandwidth $B$ as
\begin{equation}
T = T_0 + T_{\text{comm}}, \quad
T_{\text{comm}}(n,B) = \underbrace{\frac{Q}{B}\vphantom{\frac{Q_o}{n B_o}}}_{\text{scale-up}}
+ \underbrace{\frac{Q_o}{n B_o}}_{\text{scale-out}}
+ \underbrace{\tau(n)\vphantom{\frac{Q_o}{n B_o}}}_{\text{tax}}.
\label{eq:s6-step}
\end{equation}
$T_0$ is the compute time, $Q$ is the bytes each GPU transfers over the scale-up fabric, $Q_o$ is the bytes each domain transfers over the scale-out fabric, $B_o$ is the per-GPU scale-out bandwidth, and $\tau(n){=}b\big(\mu{+}\sigma(\gamma{+}\ln\frac{n}{m})\big)$ per Eq.~\ref{eq:sync-tax}. As the scale-up domain size ($n$) grows, its scale-out traffic can be spread out over more NICs, but the synchronization delay also increases (\S\ref{sec:evt}). Note that Eq.~\ref{eq:s6-step} adapts the Hockney model of collective communication (\S\ref{sec:background}), but we omit $\alpha$ costs here for clarity (but evaluate with it). Unlike \S\ref{sec:tax-model}, where $b$ is the standard deviation of a single compute block, here $b = \sum_c b_c$ is the sum of the per-block standard deviations $b_c$ over all compute blocks in a step. Today's baseline is the same model with $\tau(n){=}0$.

\myparab{Communication budget.} An architect seeks to ensure that communication is at most $\epsilon$ fraction of step time, \ie $T_{\text{comm}}/T \le \epsilon$. Substituting Eq.~\ref{eq:s6-step}, we obtain $T_{\epsilon} \equiv \tfrac{\epsilon}{1-\epsilon}T_0 \ge T_{\text{comm}}$. We now define $F(n) = \frac{Q_o}{nB_o} + \tau(n)$ as the communication \emph{floor}, \ie the part of $T_{\text{comm}}$ that is independent of scale-up bandwidth. Following Eq.~\ref{eq:s6-step}, $Q/B + F(n) \le T_{\epsilon}$, so we solve for the required bandwidth:
\begin{equation}
B_{\epsilon} \ge \frac{Q}{T_{\epsilon} - F(n)}.
\label{eq:opt-bw}
\end{equation}

\myparab{Synchronization tax.} Accounting for the synchronization delay $\tau$ increases $F(n)$, thereby increasing the required bandwidth $B_{\epsilon}$ to achieve a target $\epsilon$. Thus, we refer to $\tau$ as the \emph{synchronization tax} because it \emph{taxes} resource requirements for scale-up domains.

\myparab{Methodology.} We simulate the Llama-3-70B workload on H200 in ASTRA-sim~\cite{won2023astra} using the analytical congestion-aware backend. The simulator replays the empirical trace's 1,129 TP collectives with their measured buffer sizes and preceding compute blocks with their measured mean durations ($a_c$). Since our Llama-3-70B workload used TP groups of 4 GPUs, we model a domain of $n$ GPUs as $n/4$ TP groups. GPUs at the same TP index form an FSDP group of $n/4$, and FSDP collectives are issued at their measured positions and overlap with computation as in the trace. Each simulated GPU still executes the same per-GPU work as the measured 8-GPU workload, \ie weak scaling. GPU interconnects have the default $1\mu{s}$ latency and run at their peak bandwidth times 0.64, the measured link efficiency, so that simulated transfer times match measured H200 kernels. Every compute block pays the domain-wide synchronization tax in Eq.~\ref{eq:sync-tax} (\S\ref{sec:fitting}). We apply it to one GPU selected at random per TP group, which is $\tau_c$ slower while the other three are $\tau_c/3$ faster, so the mean compute stays $a_c$, whereas the baseline fixes every GPU at $a_c$.

We use our H200 measurements to calibrate simulations of larger scale-up domains that we could not measure directly (\eg NVL72). In these domains, training combines TP with context parallelism (CP) for long-context workloads or expert parallelism (EP) for MoE models. Since the scale-up domain is typically used for communication that lies on the critical path, the domain-wide synchronization tax propagates within two collectives, once within the TP group and subsequently across TP groups (via EP or CP). Our empirical results demonstrate that even FSDP, which is often overlapped with computation, couples the synchronization tax across all TP groups in the domain (\S\ref{sec:fitting}). This cross-group coupling motivates using the domain size $n$ in $\tau_c(n)$ while holding the TP group size fixed.

\begin{figure}[h]
\centering
\includegraphics[width=\linewidth]{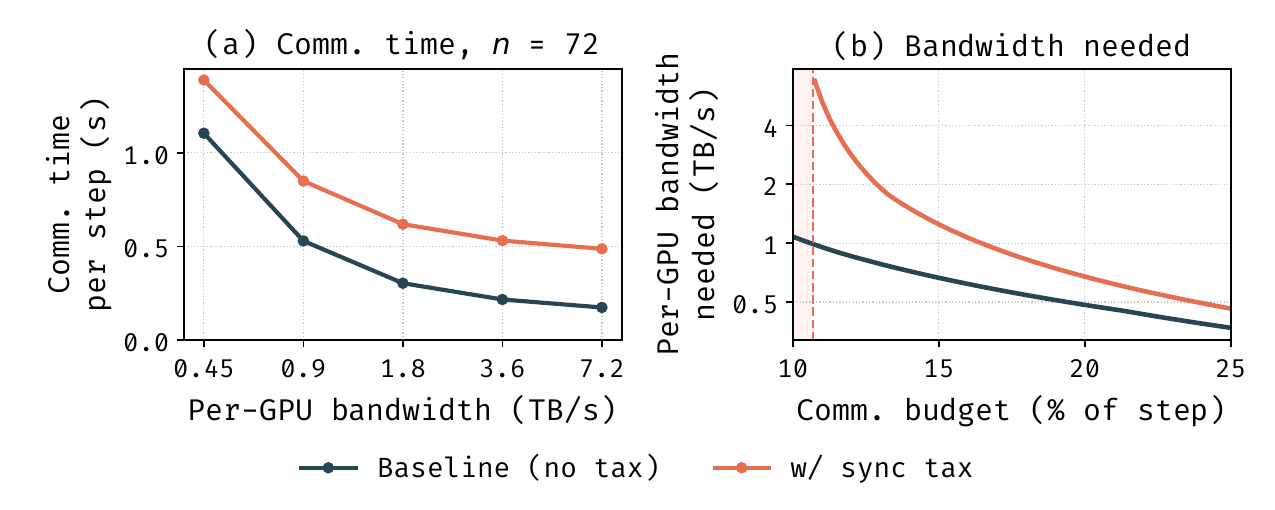}
\vspace{-3mm}
\caption{(a) When modeling the tax, communication time increases by 1.3-2.8$\times$ at $n{=}72$ with ASTRA-sim simulations for the Llama-3 70B, H200 workload. (b) Per-GPU bandwidth needed to keep communication within a budget $\epsilon$ of step time. Beyond the dashed line, no bandwidth is effective, as the tax completely dominates.}
\label{fig:bw-floor}
\end{figure}

\myparab{Results.} In Figure~\ref{fig:bw-floor}, we fix $n{=}72$ and sweep per-GPU bandwidth from 0.45--7.2~TB/s. Figure~\ref{fig:bw-floor}a shows that for 72-GPU scale-up domains, \eg NVL72, increasing per-GPU bandwidth faces diminishing returns, and increasingly so with the synchronization tax. Since adding bandwidth cannot reduce the tax, satisfying the communication budget requires shrinking transfer time, by provisioning more bandwidth than tax-free models predict. Figure~\ref{fig:bw-floor}b shows that as $\epsilon$, the target communication fraction, decreases, the required bandwidth grows under both today's provisioning model and our model with the tax. However, while the baseline predicts a gradual increase, ours requires a rapid increase to offset the tax, \ie $3.3\times$ more bandwidth at $\epsilon{=}$12\% for a 72-GPU domain. Some values of $\epsilon$ cannot be achieved because the tax is too large.

\subsection{Dividing a switch's capacity}
The synchronization tax also shifts the optimal size of scale-up domains. Architects size scale-up domains in two steps. The workload's model parallelism (TP, EP, CP) sets a minimum domain size to ensure that all blocking model-parallel collectives execute over the high-bandwidth scale-up fabric~\cite{narayanan2021efficient,zhao2025insights,chu2025scaling}. Above that, architects choose how many model-parallel groups share a switch~\cite{shou2025infinitehbd, tithi2025scaling, isaev2023calculon}, \ie how to divide the scale-up switch's capacity among GPUs. This decision navigates a fundamental trade-off: expanding a scale-up domain enables packing more GPUs within its high-bandwidth fabric, but reduces per-GPU bandwidth by increasing the number of GPUs sharing its capacity. UALink x1/x2/x4 links~\cite{ualink} and Ethernet scale-up configurations make this very choice on one chip while NVLink racks~\cite{nvlink} fix it at design time across many switch chips.

\myparab{Model.} Each scale-up switch has fixed capacity $C$ shared equally by $n$ GPUs in the domain, \ie each GPU's bandwidth is $C/n$. As in NVL72, where each GPU links to all 18 switch chips~\cite{nvlink}, we treat the switch chips of a domain as one non-blocking switch of aggregate capacity $C$. Growing $n$ partitions the same GPUs into fewer, but larger scale-up domains. This increases the $\frac{nQ}{C}$ data-movement time within the scale-up domain due to lower per-GPU bandwidth, reduces the $\frac{Q_o}{nB_o}$ scale-out time by spreading traffic across more NICs, and raises the synchronization tax $\tau(n)$ by forcing more GPUs to synchronize at each non-overlapped scale-up collective. Alg.~\ref{alg:provision} shows how this cost model is minimized to derive the optimal number of GPUs in the scale-up domain $n^*$ and per-GPU bandwidth $B^* = C/n^*$, and that the tax pushes the optimum towards \emph{fewer} GPUs per domain than the tax-free baseline (Line 4). Thus, packing more GPUs into the high-bandwidth scale-up fabric is only worth it if the reduction in scale-out communication exceeds both the increase in scale-up communication due to lower per-GPU bandwidth and a higher synchronization tax. While we exclude the tax in scale-out communication for clarity because data-parallel replicas synchronize only once per step, we still evaluate its impacts below.

\begin{algorithm}[t]
\caption{Sizing a scale-up domain with switch capacity $C$}
\label{alg:provision}
\begin{algorithmic}[1]
\Input $C$; scale-up bytes per GPU $Q{=}\sum_l q_l S_l$;
scale-out bytes $Q_o$ and per-GPU bandwidth $B_o$;
per-block compute std.\ dev.\ $b_c$; Gumbel parameters $\mu$, $\sigma$ (at fitted $m$)
\Output Optimal domain size $n^*$, per-GPU bandwidth $B^*$

\State $T_{\text{step}}(n)
    = T_0 + \frac{nQ}{C} + \frac{Q_o}{nB_o} + \tau(n)$
    \Comment{$B=C/n$, $p\alpha\approx 0$}
\State $\tau(n) = b\big(\mu + \sigma(\gamma + \ln\frac{n}{m})\big)$, $b = \sum_c b_c$ \Comment{Eq.~\ref{eq:sync-tax}}
\State Minimize $T_{\text{step}}(n)$: \Comment{$\mathrm{d}T_{\text{step}}/\mathrm{d}n = 0$}
\Statex \hspace{\algorithmicindent}$\displaystyle \frac{Q_o}{n^2 B_o} = \frac{Q}{C} + \frac{\sigma b}{n}$
    \Comment{saving $=$ cost $+$ tax}
\State $n^* \gets \tfrac{C}{2Q}\big(\sqrt{(\sigma b)^2 + \tfrac{4QQ_o}{CB_o}} - \sigma b\big)$
\State \Return $n^*$, $B^* =C/n^*$
\end{algorithmic}
\end{algorithm}

\begin{figure*}[t]
\centering
\includegraphics[width=0.75\linewidth]{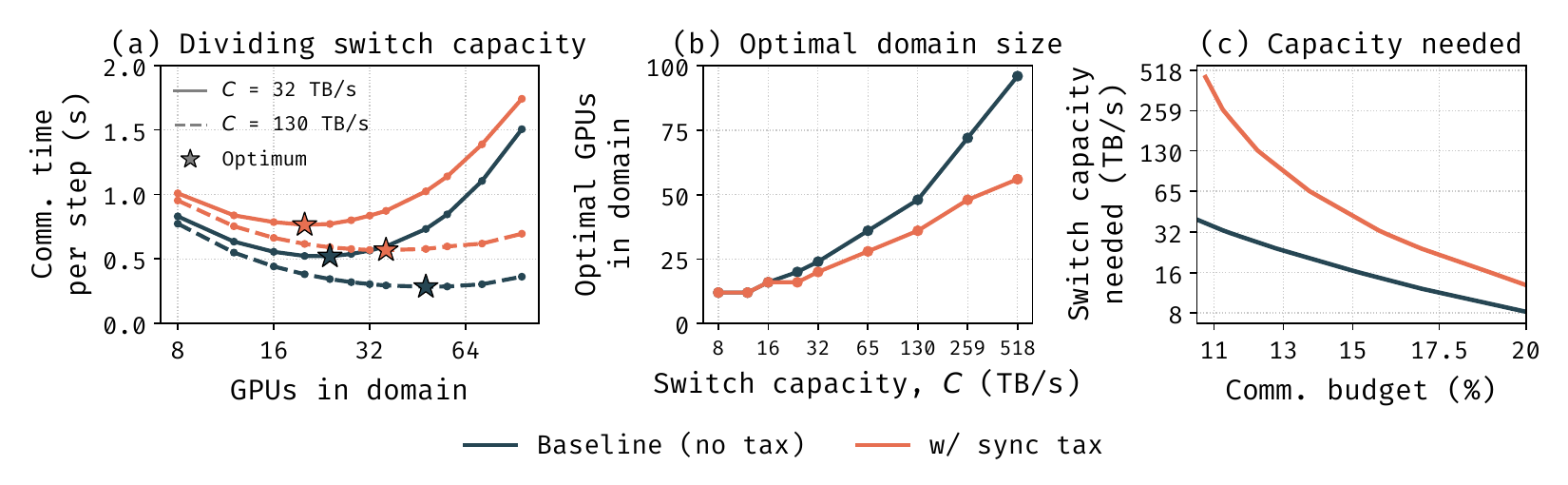}
\vspace{-2mm}
\caption{Dividing a fixed switch capacity $C$ among $n$ GPUs (Llama-3 70B, H200,
ASTRA-sim). (a) Communication time vs. domain size at 32 and 130~TB/s, where stars denote the
optimal values. (b) Modeling the tax leads to scale-up domains that should pack \emph{fewer} GPUs with more bandwidth each, and increasingly so as $C$ increases. (c) Switch capacity needed to keep communication within $\epsilon$ budget of step time.}
\label{fig:bw-switch}
\end{figure*}

\myparab{Methodology.} As in \S\ref{subsec:budget}, we evaluate our approach in ASTRA-sim on Llama-3 70B (BF16). We assume one 400~Gb/s NIC per GPU and simulate switch capacities $C$ from 8 to 518~TB/s, where a 72-GPU domain (\eg NVL72) at H200 NVLink bandwidth (450 GB/s) provides 32.4~TB/s. We divide $C$ equally among $n$ GPUs ($B{=}C/n$), sweep $n$ from 8 to 144, and report the best simulated domain size ($n$) at each capacity ($C$). For analyses with the $\epsilon$ budget, we take the minimum achieved communication share over $n$ at each $C$. Note that one NVLink~5 switch chip provides 3.6~TB/s per direction.

\myparab{Results.} Figure~\ref{fig:bw-switch}a shows that for both $C{=}32$ TB/s and $C{=}130$ TB/s, the tax raises the communication time at every $n$ and pushes the optimal scale-up domain size toward fewer GPUs. Figure~\ref{fig:bw-switch}b shows that this shift increases with capacity: after accounting for the tax, the optimal scale-up domain at 518~TB/s is 56 instead of 96 GPUs, \ie 17--42\% fewer GPUs with 20--71\% more bandwidth each. Including the tax in scale-out communication (8--13~ms vs. 181--356~ms in scale-up) does not change these optima. Although communication time is relatively flat near the optimum, our model and the tax-free baseline diverge significantly in how they rank designs across the whole curve. The tax-free baseline overestimates the cost of shrinking domain size, and it rules out several designs that are within 10\% of optimal. Even for its optimal design, the baseline underpredicts its communication time by 1.5-3$\times$ by neglecting the tax. Figure~\ref{fig:bw-switch}c shows that after accounting for the tax, meeting a communication budget of $\epsilon{=}12$\% of step time requires a switch capacity of 156 rather than 28~TB/s ($5.6\times$, \ie 44 vs. 8 chips). The required capacity grows far more steeply with the tax than without it as the budget shrinks, a gap of $1.6\times$ at $\epsilon{=}$20\% to $10\times$ at $\epsilon{=}$11\%. Increasing switch capacity from 32 to 518 TB/s only reduces communication time by 37\%, not the 68\% reduction promised by the tax-free simulation. Therefore, it is more worthwhile to take steps to reduce the tax than to increase bandwidth under tight communication budgets.

\myparab{Wider model parallelism.}  Wider TP could narrow the gap between our model and the baseline by potentially leaving less room to shrink domain size. We repeat the sweep with TP degrees of TP=8 and 16, keeping per-GPU compute fixed while increasing TP data volume by $2\times$ for TP=8 and $4\times$ for TP=16 and reducing FSDP volume in proportion. While both models choose the same optimal values with TP=16 at 32 and 130~TB/s, at larger $C$ the tax still favors fewer GPUs with more bandwidth each compared to the baseline (TP=8: 24 vs. 32 GPUs at 130~TB/s, 48 vs. 64 at 518~TB/s).

\subsection{Jointly scaling GPU FLOPS}
We now jointly scale compute FLOPS, scale-up bandwidth, and scale-out bandwidth. The tax's impact depends on how the spread $b$ scales relative to compute $a$, \ie how the ratio $r{=}b/a$ changes.

\myparab{Methodology.} We scale FLOPS by a factor $f{\in}\{2,4,8\}$ over H200 BF16 FLOPS by dividing compute runtime by $f$ and multiplying switch capacity ($C{=}f{\cdot}32.4$~TB/s), scale-out bandwidth ($B_o{=}f{\cdot}50$~GB/s), and memory bandwidth by $f$. We consider how $r$ changes by scaling per-block spread as $b_c/f^{\kappa}$. $\kappa{=}1$ maintains a constant $r$ (\ie both $a_c$ and $b_c$ reduce by $f$), and $\kappa{\approx}0.3$ increases $r$ (\ie $b_c$ reduces by less than $f$), matching our measured change in $r$ for Llama-3 70B from Ampere to Hopper ($b{\approx}a^{0.34}$). At each $f$, we report the best communication share over $n$ (8--144 GPUs).

\begin{figure}[h]
\centering
\includegraphics[width=\linewidth]{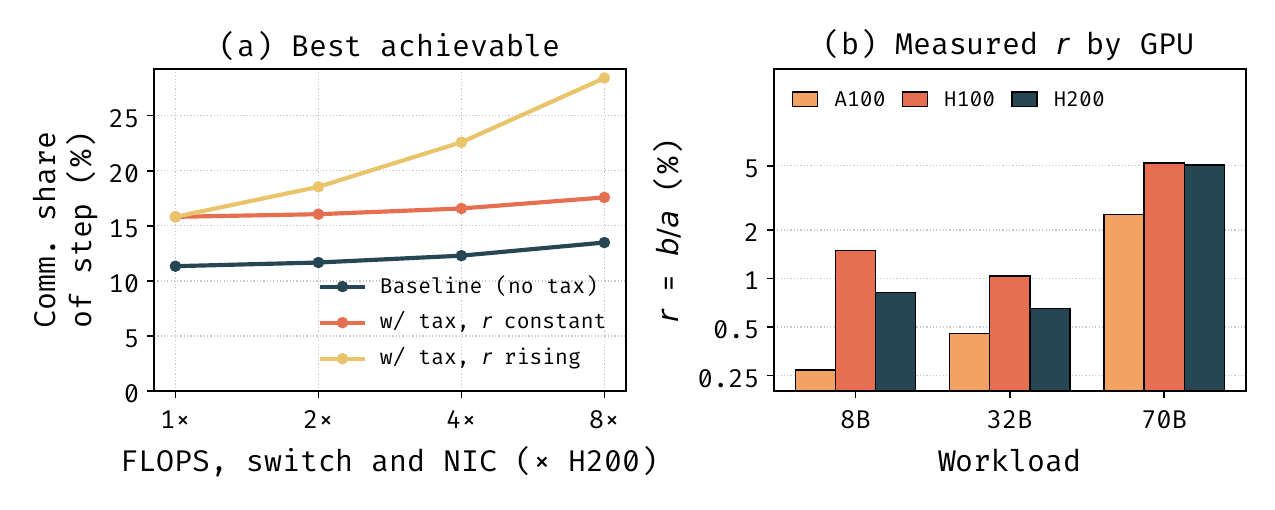}
\vspace{-2mm}
\caption{Jointly scaling bandwidth with FLOPS. Flat $r$ means compute variation scales with FLOPS; rising $r$ means it grows sublinearly.}
\label{fig:bw-generations}
\end{figure}

\myparab{Results.} Figure~\ref{fig:bw-generations}a shows how the fraction of end-to-end time spent on communication changes as $f$ scales. Incorporating the tax naturally shifts the curve upwards due to synchronization overheads. While both the tax-free baseline and our model with \emph{constant} $r$ are relatively stable as $f$ scales, only slightly increasing because link latency starts to dominate at large $f$, our model with \emph{increasing} $r$ (as reflected in the empirical data, per Figure~\ref{fig:bw-generations}b) shows communication's share of end-to-end time increasing much more rapidly with $f$, \ie $2.1\times$ higher than the tax-free baseline at $f{=}8\times$. If variation keeps lagging compute as Figure~\ref{fig:bw-generations}b indicates, the tax's influence over scale-up provisioning grows with each generation.

\section{Related Work}
\label{sec:related}

\myparab{Scale-up architecture.} Recent works propose more efficient scale-up designs, \eg NVLink racks~\cite{ishii2022nvlink,nvlink}, optical fabrics~\cite{googletpuv4, shou2025infinitehbd, khani2021sip}, all-to-all supernodes~\cite{zuo2025serving,liao2025ub}, and wafer-scale fabrics~\cite{rashidi2025fred, wang2026watos, lie2023cerebras}. Other works co-optimize FLOPS, bandwidth, and parallelism~\cite{isaev2023calculon,tithi2025scaling, won2024libra} but assume that GPUs arrive at collectives simultaneously.

\myparab{Modeling and simulation.} Most analytical models~\cite{isaev2023calculon,hsia2024mad,tithi2025scaling,won2024libra}, profile-driven simulators~\cite{bang2024vtrain,li2025triosim,liang2025lumos, chakra}, multi-GPU simulators~\cite{won2023astra,wang2025simai}, and cycle-accurate simulators~\cite{khairy2020accel} assume deterministic performance. Recent work acknowledges runtime non-determinism and averages samples to obtain more accurate estimates for profile-driven simulators~\cite{chung2025swift}. PRISM~\cite{golden2025prism} enables non-deterministic simulation through Monte Carlo sampling, but it does not investigate or model this variation (or synchronization itself).

\myparab{ML synchronization.} Prior work studies variability in HPC and earlier ML workloads~\cite{hoefler2010characterizing, shen2024llamp,sinha2022not, go2025characterizing, kokolis2025revisiting, petrini2003case, mogilevsky2026quantifying, jain2024pal}, but does not model synchronization delay. Recent works highlight synchronization challenges in \emph{scale-out} networks~\cite{lin2025understanding,jiang2024megascale,wu2024falcon,warraich2025optireduce,deng2025minder,yao2025holmes, lao2024trainmover,dean2013tail}. Prior work studies DVFS as a control knob~\cite{chung2024reducing,stojkovic2025dynamollm, wang2025using, kakolyris2025throttll} and as a source of persistent variation~\cite{sinha2022not,go2025characterizing,kurzynski2025lit}, whereas other works suggest delays are driven by CPU kernel launches~\cite{aimuyo2026flashmoe,cheng2026mpk,trifan2025eliminating}, work imbalance~\cite{lin2025understanding}, or garbage collection~\cite{jiang2024megascale}. In contrast, our attribution algorithm (\S\ref{sec:attribution}) identifies \emph{dynamic} variation in GEMM kernels as a key source of the tax \emph{at the collective}. Recent work studies specific persistent scale-up stragglers on AMD GPUs that can be partially mitigated through one-time power capping~\cite{kurzynski2025lit}; we do not observe a similar deterministic pattern in NVIDIA GPUs and thus model it as a fundamental consequence of distributed execution.

\myparab{Trace analysis.} Several works localize severe slowdowns in ML jobs~\cite{deng2025mycroft,yao2025holmes, cui2025xputimer,li2025malleus}, \eg scale-out network faults, or detect stragglers using heuristic anomaly rankings~\cite{lin2025understanding,guan2025perftracker}. PyTorch's HTA~\cite{bhatnagar2022hta} analyzes each rank individually and requires injecting CUDA synchronization events that perturb behavior. In contrast to iteration-level path recovery~\cite{bhatnagar2022hta, liang2025lumos, lin2025understanding}, we must recover critical paths \emph{for each collective} to accurately attribute the synchronization tax.

\myparab{Collective communication.} Recent work has explored topology-specific collective algorithms~\cite{cai2021synthesizing,taccl,won2024tacos,zhao2024forestcoll,liu2024rethinking,cao2025syccl,huang2021communication,rashidi2022themis}, novel abstractions and libraries~\cite{cowan2023mscclang,hwang2026msccl,si2025collective,hidayetoglu2025hiccl,liu2025resccl}, in-network aggregation~\cite{klenk2020network, sapio2021scaling,lao2021atp}, and parallelism strategies~\cite{zheng2022alpa,nam2025meshslice, chu2025scaling}. Other works study opportunities for compute-communication overlap~\cite{wang2022overlap,chen2024centauri,pati2024t3, rashidi2021enabling, lee2025characterizing, devraj2025efficient,devraj2026breaking}, also present in our analysis (\S\ref{sec:bw-model}).

\section{Conclusion}
\label{sec:conclusion}
While hardware simulators model communication solely as data movement, our measurements of LLM workloads point to a neglected overhead: synchronization across GPUs. By understanding and modeling this overhead, we show how synchronization delays shift the optimal size and bandwidth of future scale-up domains.

\bibliographystyle{ACM-Reference-Format}
\bibliography{reference}
\clearpage %
\appendix
\newpage
\section{FSDP and per collective results}

\label{sec:appendix-measurements}

\begin{figure}[h]
\centering
\includegraphics[width=0.8\linewidth]{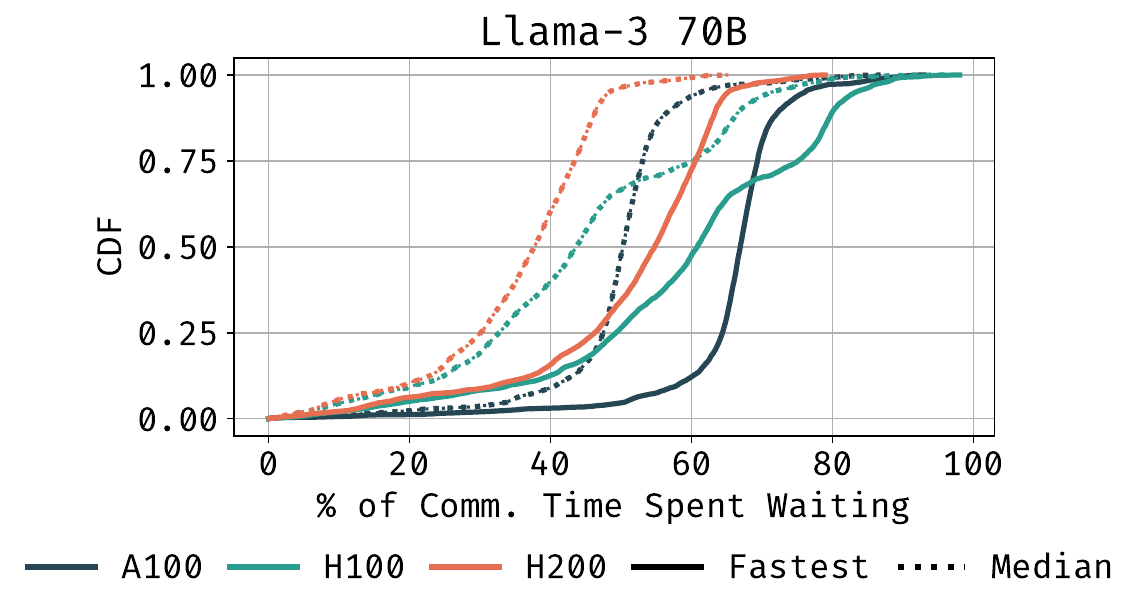}
\caption{CDFs of the synchronization tax---shown as the percent of NCCL collective communication time that the fastest and median ranks spend waiting---across FSDP collective instances for all workloads and architectures.}
\label{fig:fsdp-cdf}
\end{figure}

In addition to our CDF-based analysis of TP communication overheads, we observe similar trends for FSDP communication events when running Llama-3 70B across multiple GPU architectures.

\begin{figure*}[h]
\centering
\includegraphics[width=0.9\linewidth]{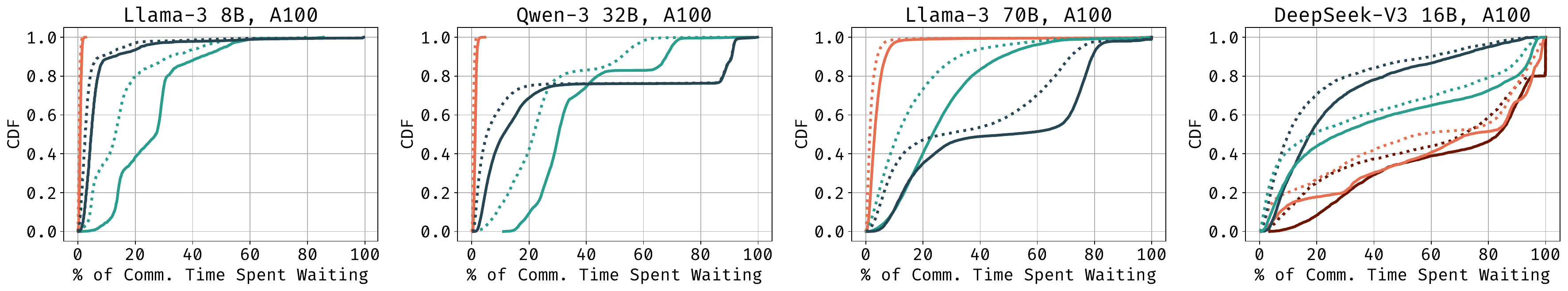}
\end{figure*}

\begin{figure*}[h]
\centering
\includegraphics[width=0.9\linewidth]{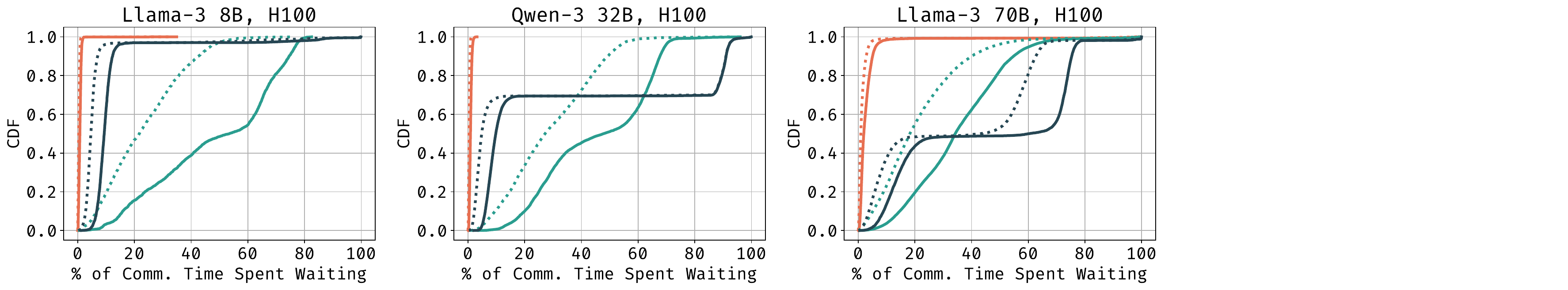}
\end{figure*}

\begin{figure*}[h]
\centering
\includegraphics[width=0.9\linewidth]{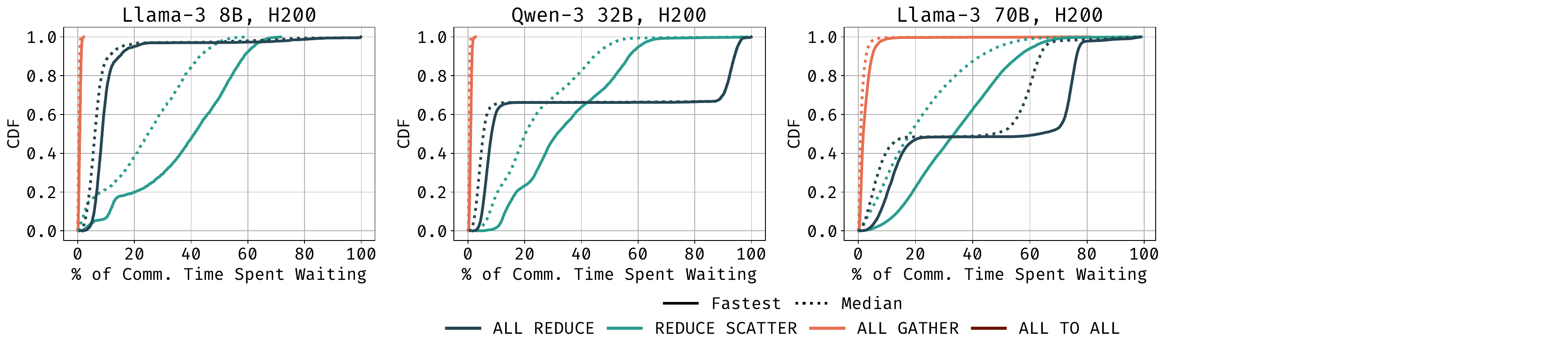}
\caption{CDFs of the synchronization tax broken down by collective and workloads across TP collectives (in addition to EP for DeepSeek) and architectures.}
\label{fig:H200-cdf}
\end{figure*}

We also analyze how the synchronization tax affects different collective communication operations, shown in Figure~\ref{fig:H200-cdf}. When using tensor parallelism, operations such as \allgather are commonly invoked in combination after a \reducescatter. This means that very few compute kernels run in between these operations, resulting in very little preceding variation. This lack of variation shows up in our analysis as \allgather experiencing less of the synchronization tax. However, it should be noted that this is due to the context the collective operation is called under. In general, the design of scale-up domains is based on collective communication that is used to exchange results from preceding computation, as we model in \S\ref{sec:tax-model} and \S\ref{sec:bw-model}.

\section{Shifting slowest GPU}
\label{sec:slowest-id}
\begin{figure*}[h]
\centering
\includegraphics[width=0.68\linewidth]{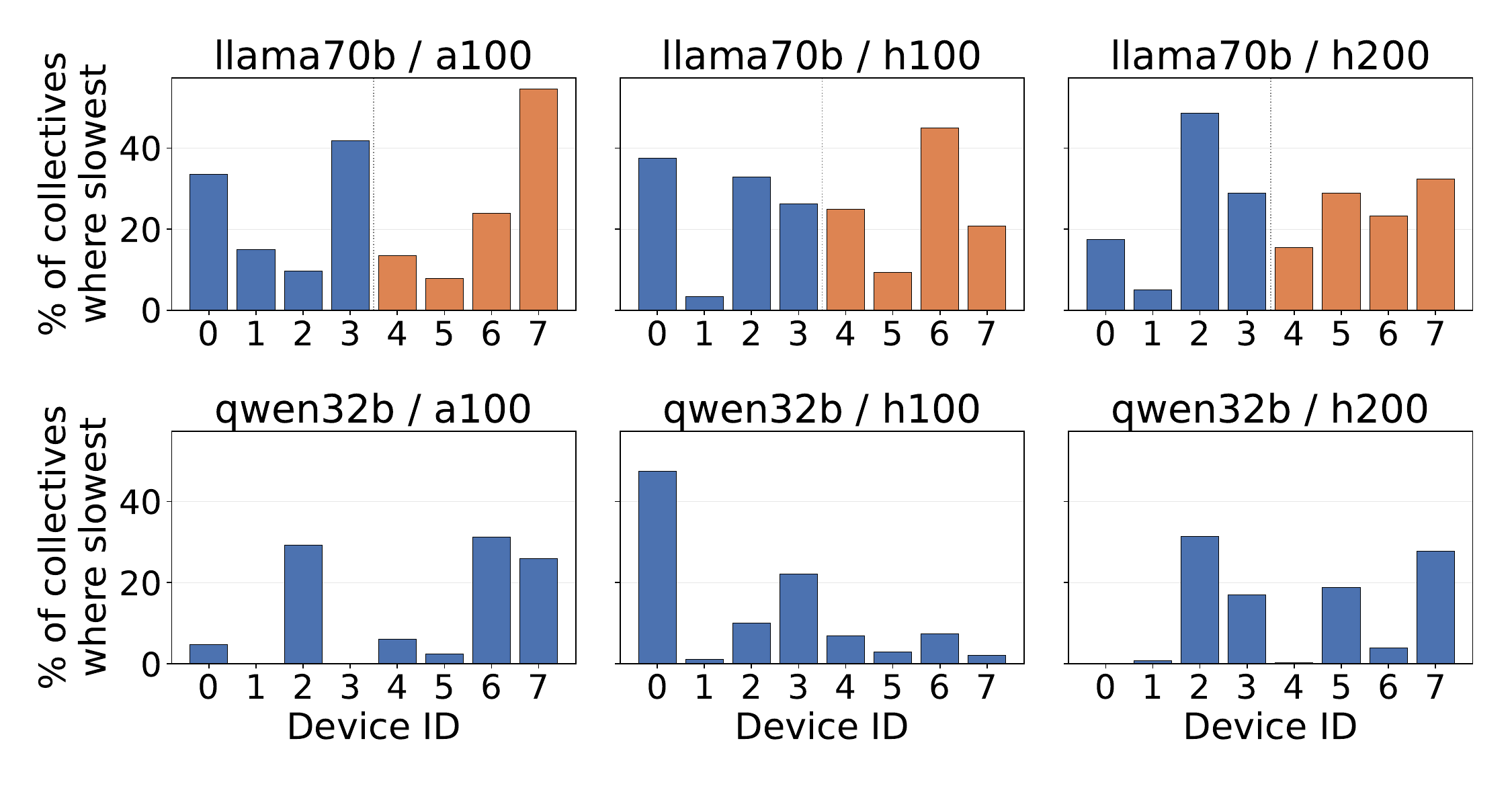}
\caption{GPU device ID of the slowest rank across all TP \reducescatter collectives. Since Llama-3 70B used two TP groups of size 4 each, we calculate the slowest rank separately for the collectives in each TP group and distinguish between these two groups by color.}
\label{fig:slowest-id}
\end{figure*}

Across the Llama-3 70B and Qwen-3 32B traces, we quantify the frequency that each GPU device was the slowest rank for all TP \reducescatter collectives. We specifically focus on \reducescatter because \allgather occurs directly after \reducescatter, which naturally has no preceding compute (and in fact features a uniform distribution across ranks). Figure~\ref{fig:slowest-id} shows that while some devices are more likely to be the slowest rank than others, only in one workload does a single GPU ever exceed 50\% of the total collectives. While the distribution is not uniform, alluding to the DVFS effects we observe in \S\ref{sec:gemm-variation} and power variation results found in~\cite{sinha2022not, kurzynski2025lit}, it is also not predictable, as rarely is any single device the slowest rank for over half of the collective instances. As a result, mitigations that rely on offline profiling and detection of the slowest rank cannot eliminate the synchronization tax due to the straggler's changing identity.

\section{Existing critical-path analysis tools}
\label{sec:eroica}
The core challenge for our critical-path analysis tool was that we needed to be able to attribute the cross-rank variation for each collective to the variation between ranks' critical paths leading up to the collective. This is challenging because (1) lightweight PyTorch profiling tools (\eg Kineto) do not encode cross-stream dependencies and (2) tracking these dependencies requires turning on PyTorch's experimental \texttt{enable\_cuda\_sync\_events} flag which distorts runtime behavior (see below). In \S\ref{sec:critical-path}, we show how we are able to circumvent these challenges using our algorithm, specifically by using a MIN-HEAP to efficiently keep track of in-flight events and infer cross-stream dependencies.

While PyTorch offers an experimental trace analysis tool~\cite{bhatnagar2022hta}, it only takes in a single rank's trace as input, requires adding extra CUDA synchronization events that distort runtime behavior (our Llama-3 70B workload fails after 30 iterations when using it), and only infers the longest path for the entire duration rather than $\mu s$ granularity before each collective. We also tried using the Chakra execution trace~\cite{chakra} that is co-designed with the ASTRA-sim simulator~\cite{won2023astra}, but we found that it only captures explicit caller-callee control flow (\eg a CPU function that calls another function) and was unable to infer dependencies between different CUDA streams on the same GPU, which is critical for our use case of per-collective critical-path tracing.

EROICA~\cite{guan2025perftracker} also identifies a critical path. However, this method uses a pre-defined ranking of importance: the event with the highest priority is always selected as part of the critical path. This summarizes where time is spent but does not trace dependencies, and excludes concurrent operations that do not align with the pre-defined ranking. Instead of tracing a sequence of dependent events, the method computes the fraction of time during which the highest-priority events are exposed. EROICA guarantees that the constructed path accounts for the full duration of execution, but does not reveal dependency structures. By ignoring execution contexts such as streams and operation boundaries, EROICA can introduce artificial transitions between unrelated events, implicitly suggesting dependencies that do not exist.

Finally, while PRISM~\cite{golden2025prism} reports cross-stream dependencies, the paper does not specify how these edges are recovered from traces, or how the DAG is constructed and traversed. PRISM focuses on sampling from kernel distributions for iteration-level simulation. In contrast, our goal is to decompose individual collectives into the transfer time and synchronization delay, which we attribute along critical paths. By imposing that critical paths be accurate at the per-collective granularity, our approach ensures $\mu$s-scale verification by comparing attributed critical-path variation to observed variation at each collective.

\clearpage
\section{Critical-path analysis algorithms}
\begin{figure*}[h]
\centering
\includegraphics[width=1.0\linewidth]{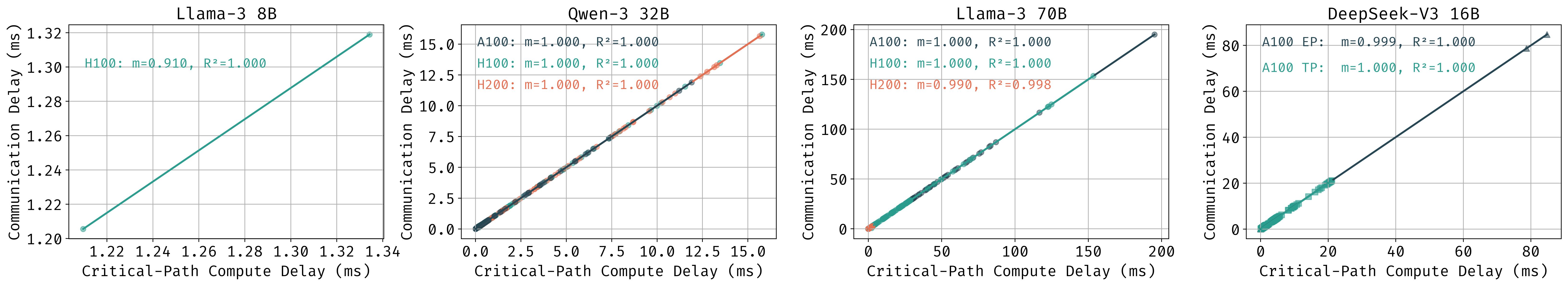}
\caption{Correlational validation of collective communication events comparing both GPU and CPU critical paths.}
\label{fig:mix-comp}
\end{figure*}
\label{sec:algo}
We describe the different types of edges we add to construct the DAG in more detail below.
\begin{compactitem}
\item \emph{In-stream.} We add an edge from the preceding event on the same CUDA stream or CPU thread to the current event, due to a possible dependency from sequential execution within a stream.

\item \emph{Cross-stream.} To discover possible cross-stream dependencies, we maintain a MIN-HEAP of in-flight events keyed by completion time. When we process a new event at time $t$, we pop all heap entries whose completion is at or before $t$ and add an edge to the current event, as it is the last event that could have delayed the current event. We then push the new event onto the heap with its completion time.

\item \emph{Launch.} We connect each CPU-side kernel launch to the GPU kernel it dispatches via its shared CUPTI correlation.

\item \emph{Synchronization.} For \texttt{cudaStreamSynchronize}, we identify the most recently completed GPU kernel and add an edge to its parent CPU event. Launch edges connect CPU to GPU while synchronization edges connect GPU to CPU.
\end{compactitem}

Alg.~\ref{alg:construction} provides the trace DAG construction algorithm, Alg.~\ref{alg:traversal} provides the traversal algorithm to find per-rank, per-collective critical paths, and Alg.~\ref{alg:analysis} uses the results of the previous algorithms to attribute and compare variation across ranks for each collective.

\begin{algorithm}[H]
\caption{BuildDAG($\mathit{timeline}$, $\mathit{comm\_stream}$)}
\begin{algorithmic}[1]
\Input{$\mathit{timeline}$: map of stream $\to$ event list; $\mathit{comm\_stream}$: communication stream name}
\Output{$G = (V, E)$: execution DAG; $C \subseteq V$: communication nodes}

\State $G \gets (\{r\}, \emptyset)$, \; $r$ dummy root with $\mathrm{id} = 0$
\State $H \gets \emptyset$ (min-heap on $t_e$); \; $\mathit{last\_fin} \gets \bot$; \; $\mathit{last\_strm}[s] \gets \bot$
\State $C \gets \emptyset$; \; $t_{\mathit{comm}} \gets 0$; \; $\mathit{id} \gets 1$; \; $\mathit{ptr}[s] \gets 0$ for all $s$

\medskip
\State Let $\mathit{active}(s) \equiv \mathit{ptr}[s] < |\mathit{timeline}[s]|$
\While{$\exists\, s : \mathit{active}(s)$}
    \State $s^* \gets$ stream with earliest next event among all $s$ with $\mathit{active}(s)$
    \State $e \gets$ next event of $s^*$; \quad $\mathit{ptr}[s^*] \mathrel{+}= 1$

    \While{$H \neq \emptyset$ \textbf{and} $H.\mathit{min}().t_e \leq e.t_s$}
        \State $\mathit{last\_fin} \gets H.\mathit{pop}().\mathit{id}$
    \EndWhile

    \State $v \gets \text{new GPU node}(\mathit{id},\, e)$;
    \State $G.V \gets G.V \cup \{v\}$
    \State $p \gets \mathit{last\_fin}$ if $\mathit{last\_fin} \neq \bot$ else $r.\mathit{id}$
    \State $G.E \gets G.E \cup \{(p,\, \mathit{id})\}$
    \If{$\mathit{last\_strm}[s^*] \neq \bot$}
        \State $G.E \gets G.E \cup \{(\mathit{last\_strm}[s^*],\, \mathit{id})\}$
    \EndIf
    \State $\mathit{last\_strm}[s^*] \gets \mathit{id}$; \quad $H.\mathit{push}(v)$

    \If{$s^* = \mathit{comm\_stream}$}
        \State $C \gets C \cup \{\mathit{id}\}$;
        \State $t_{\mathit{comm}} \gets \max(t_{\mathit{comm}},\; e.t_e)$
    \EndIf
    \If{$e.\mathit{corr} \in \mathit{timeline}[\texttt{driver}]$}
        \State $\ell \gets \text{new CPU node}(\mathit{id}{+}1,\;\texttt{cudaLaunchKernel})$
        \State $G.V \gets G.V \cup \{\ell\}$;
        \State $G.E \gets G.E \cup \{(\ell.\mathit{id},\, \mathit{id})\}$
        \State $\mathit{id} \gets \mathit{id} + 1$
    \EndIf
    \State $\mathit{id} \gets \mathit{id} + 1$
\EndWhile
\State \Return $(G,\; C)$
\end{algorithmic}
\label{alg:construction}
\end{algorithm}

\begin{algorithm}[H]
\caption{Traverse($G$, $\mathit{node\_id}$, $\mathit{start\_id}$, $C$)}
\begin{algorithmic}[1]

\State $\mathit{comp} \gets 0$; \quad $\mathit{path} \gets \{\}$;

\While{true}
    \State $v \gets G.\mathit{get}(\mathit{node\_id})$

    \If{$(\mathit{node\_id} \neq \mathit{start\_id} \land \mathit{node\_id} \in C)$ \textbf{or} $\mathit{node\_id} = 0$}
        \State \Return $\mathit{comp}, \mathit{node\_id}, \mathit{path}$
    \EndIf

    \State $P \gets G.\mathit{parents}(\mathit{node\_id})$
    \If{$P = \emptyset$}
        \State \Return $\mathit{comp}, \mathit{node\_id}, \mathit{path}$
    \EndIf

    \State $p^* \gets \arg\max_{p \in P} \; p.t_e$
    \State $p_s \gets$ same-stream parent in $P$ if exists

    \If{$p_s \neq \bot \land |p^*.t_e - p_s.t_e| < \delta$}
        \State $p^* \gets p_s$
    \EndIf

    \If{$\mathit{node\_id} \neq \mathit{start\_id}$}
        \State $\mathit{comp} \mathrel{+}= v.d$; \quad $\mathit{path}[v.\mathit{name}] \mathrel{+}= v.d$
    \EndIf

    \State $\mathit{node\_id} \gets p^*.\mathit{id}$
\EndWhile
\end{algorithmic}
\label{alg:traversal}
\end{algorithm}

\begin{algorithm}[H]
\caption{AnalyzeGroup($\mathit{group}$, $\mathit{comms}$, $\mathit{comps}$,\\$\mathit{ancs}$, $\mathit{paths}$)}
\begin{algorithmic}[1]

\For{each event $j$}
    \State $a^* \gets \min_{r \in \mathit{group}} \; \mathit{ancs}[r][j]$

    \For{each $r \in \mathit{group}$}
        \State $\mathit{comp}[r] \gets \mathit{comps}[r][j]$
        \State $\mathit{comm}[r] \gets \mathit{comms}[r][j]$
        \State $\mathit{dur}[r] \gets \mathit{paths}[r][j]$

        \State $k \gets \mathit{ancs}[r][j]$
        \While{$k \ge 0$ \textbf{and} $\mathit{ancs}[r][k] > a^*$}
            \State $\mathit{comp}[r] \mathrel{+}= \mathit{comps}[r][k] + \mathit{comms}[r][k]$
            \State merge $\mathit{paths}[r][k]$ into $\mathit{dur}[r]$
            \State $k \gets \mathit{ancs}[r][k]$
        \EndWhile

        \If{$k \ge 0$}
            \State $\mathit{comp}[r] \mathrel{+}= \mathit{comps}[r][k] + \mathit{comms}[r][k]$
        \EndIf
    \EndFor
\EndFor
\end{algorithmic}
\label{alg:analysis}
\end{algorithm}

When comparing a GPU critical path directly to a CPU path, the difference in profiling method requires our algorithm to make adjustments to the GPU timeline to account for gaps between consecutive kernels. These gaps can be caused by GPU scheduling overhead that is not captured by the profiler. In contrast, the CPU profiler accounts for all timesteps during execution, causing a discrepancy when directly comparing these timelines with each other. These comparisons only account for <5\% of our total collective events, as most critical paths lie entirely on the GPU (>90\%). To ensure that these GPU timeline adjustments do not falsely validate our algorithm, we separate these events and evaluate them in Figure~\ref{fig:mix-comp}. We do not make any adjustments when comparing two GPU paths or two CPU paths to each other. The scatterplots in \S\ref{sec:attribution} therefore show variation measured directly from event executions, rather than a quantity that reduces to the barrier-to-barrier interval by
construction. Regardless, delay injection experiments (specifically, the CPU-side injections) confirm our algorithm's ability to recover critical paths even when different ranks have critical paths that differ between GPU and CPU across ranks.

\section{Controls for critical-path analysis}
\label{sec:neg-controls}
\begin{figure*}[t]
\centering
\begin{subfigure}[b]{0.3\textwidth}
 \includegraphics[width=\linewidth]{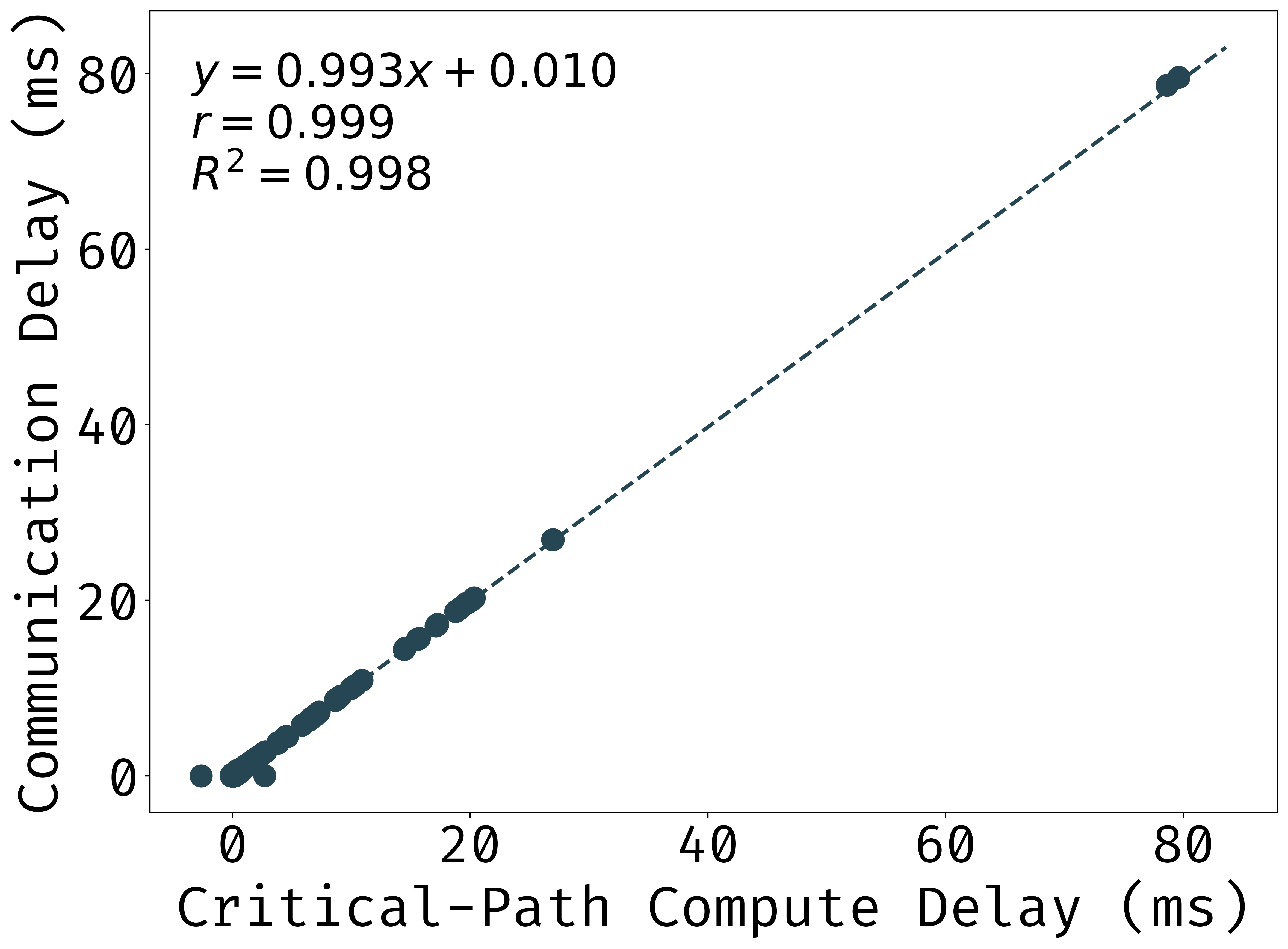}
 \caption{Our algorithm.}
 \label{subfig:og}
\end{subfigure}
\centering
\begin{subfigure}[b]{0.3\textwidth}
 \includegraphics[width=\linewidth]{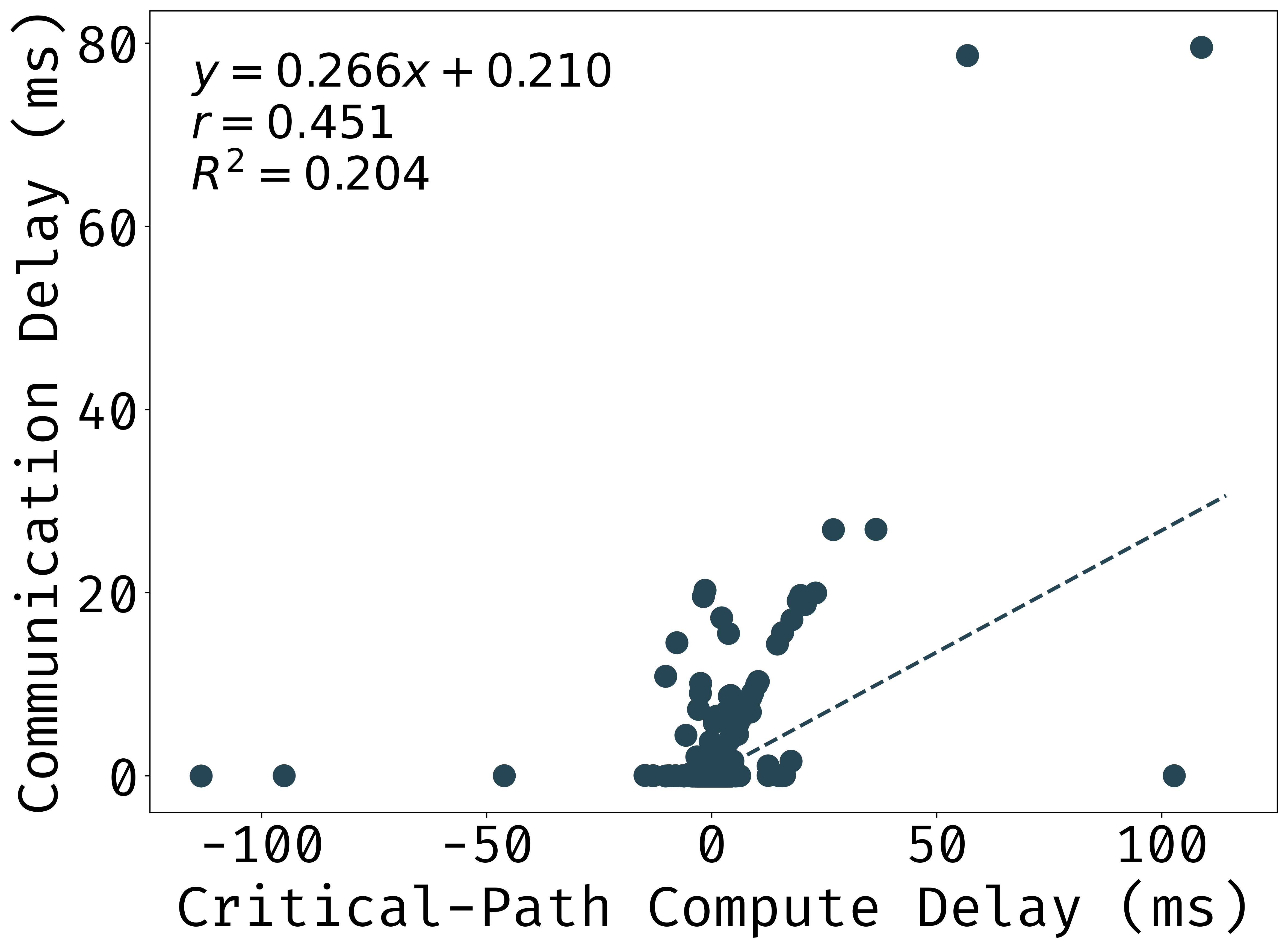}
 \caption{Longest-parent heuristic.}
 \label{subfig:longest-parent}
\end{subfigure}
\centering
\begin{subfigure}[b]{0.31\textwidth}
 \includegraphics[width=\linewidth]{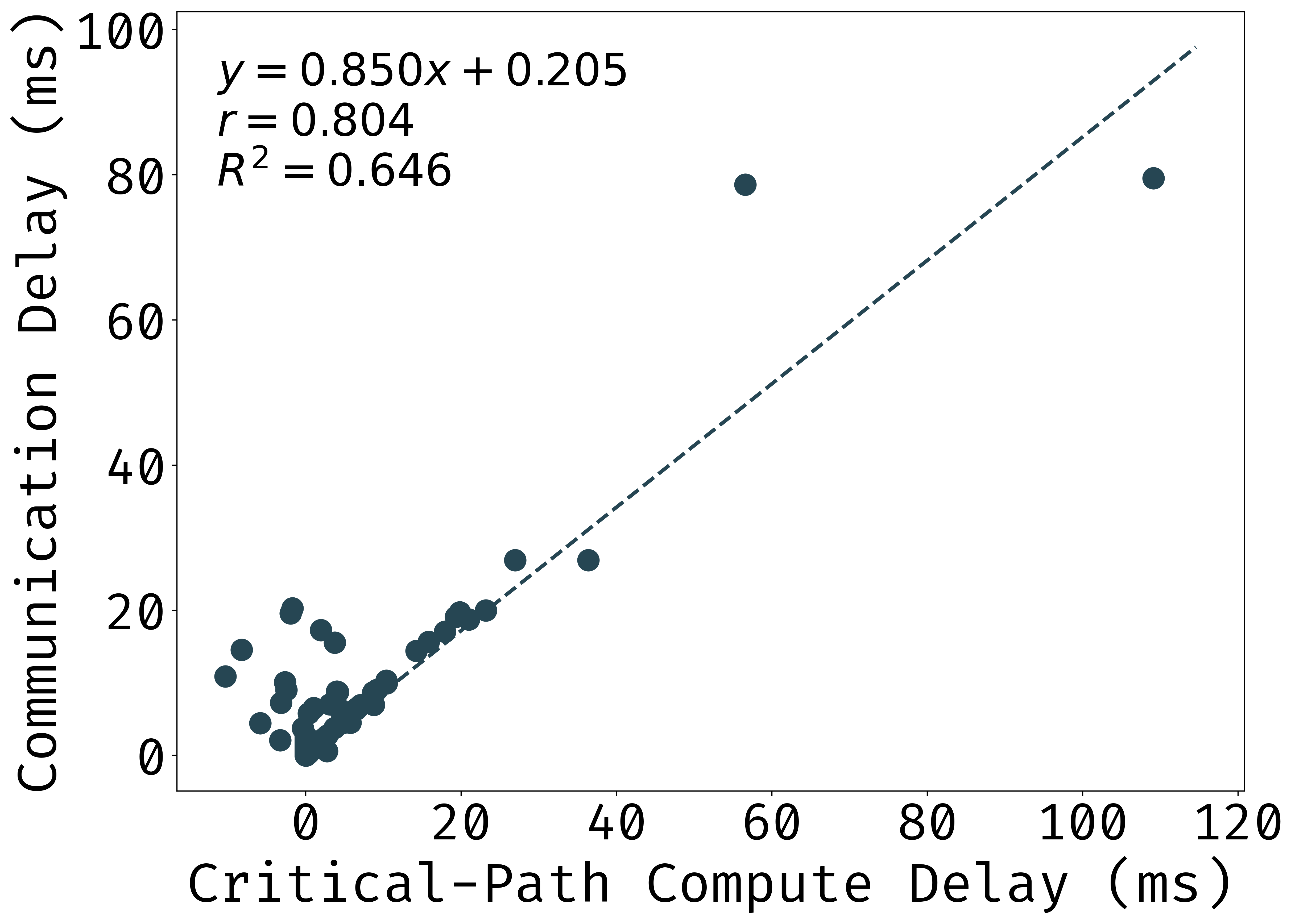}
 \caption{Without inferred cross-stream edges.}
 \label{subfig:cross-stream}
\end{subfigure}
\caption{Comparison of our algorithm (\ref{subfig:og}) to negative controls that either (1) traverse upwards via the parent with the longest duration (\ref{subfig:longest-parent}) or (2) construct a graph without inferring cross-stream edges (\ref{subfig:cross-stream}) on the Llama-3 70B-H200 workload.}
\label{fig:neg-controls}
\end{figure*}

We compare our critical-path finding algorithm (\S\ref{sec:attribution}, \S\ref{sec:algo}) to two negative controls for the Llama-3 70B-H200 workload. Figure~\ref{fig:neg-controls} demonstrates that naive approaches cannot recover the synchronization tax from compute events in the traces, and that our linear fits are the result of the algorithm's recovered paths, not just accounting for time between consecutive collectives. First, we implement a baseline that takes the parent with the longest duration instead of the most recent parent. Figure~\ref{subfig:longest-parent} shows that this baseline cannot identify the sequence of events that resulted in the synchronization delay observed at the collective, with $R^{2}=0.204$ compared to our algorithm's $R^{2}=0.998$.

We implement another baseline that uses the same traversal algorithm as ours, but does not add edges for \emph{inferred} cross-stream dependencies. Recall from \S\ref{sec:attribution} that our algorithm maintains a MIN-HEAP of in-flight events keyed by completion time to infer likely cross-stream dependencies during DAG construction itself. Figure~\ref{subfig:cross-stream} shows that this baseline is also unsuccessful at finding the right subset of preceding events that culminated in the synchronization delay, with $R^{2}=0.646$. These results highlight that recovering the critical path before each collective is neither trivial nor a self-fulfilling property of the barrier-to-barrier traversal.

\section{Compute blocks}
\label{sec:comp-blocks}
We provide an example of compute blocks by enumerating the index-to-compute-block mapping for the Llama-3 70B workload below. Each compute block is an ordered sequence of GEMM kernels that executes between consecutive collectives, as referenced in Figure~\ref{fig:gumbel-val}. We have appended the $M$, $K$, and $N$ GEMM kernel dimensions to the end of each kernel name.

{\scriptsize
\smallskip\noindent\textbf{0:}
\begin{packeditemize}
  \item \path{nvjet_tst_192x192_64x4_2x1_v_bz_coopB_TNN_16384_7168_8192}
  \item \path{nvjet_tst_192x192_64x4_2x1_v_bz_coopB_TNN_16384_8192_7168}
  \item \path{nvjet_tst_192x192_64x4_2x1_v_bz_coopB_TNN_16384_8192_7168}
\end{packeditemize}

\smallskip\noindent\textbf{1:}
\begin{packeditemize}
  \item \path{nvjet_tst_192x192_64x3_2x1_v_bz_coopB_NNN_16384_7168_8192}
  \item \path{nvjet_tst_128x256_64x4_2x1_v_bz_coopA_NTN_7168_16384_8192}
  \item \path{nvjet_tst_192x192_64x3_2x1_v_bz_coopB_NNN_16384_7168_8192}
  \item \path{nvjet_tst_128x256_64x4_2x1_v_bz_coopA_NTN_7168_16384_8192}
  \item \path{nvjet_tst_192x192_64x3_2x1_v_bz_coopB_NNN_16384_8192_7168}
  \item \path{nvjet_tst_128x256_64x4_2x1_v_bz_coopA_NTN_8192_16384_7168}
\end{packeditemize}

\smallskip\noindent\textbf{2:}
\begin{packeditemize}
  \item \path{nvjet_tst_192x192_64x3_2x1_v_bz_coopB_NNN_16384_32064_8192}
  \item \path{nvjet_tst_192x192_64x3_1x2_h_bz_coopB_NTN_32064_16384_8192}
\end{packeditemize}
}

\section{Fitting the GEVD}
\label{sec:frechet}
\myparab{Setup.} Consider a collective with $n$ participating ranks. Let $X_1,\ldots, X_n$ denote the critical-path computation time on each rank (\S\ref{sec:attribution}), which we treat as approximately i.i.d. draws from a distribution $F$. This approximation is justified by the fact that all ranks execute the same kernels on identical hardware. While we occasionally observe dependencies between ranks (\eg NUMA node placement), we note that EVT does not strictly require i.i.d. samples and tolerates both weak dependence across ranks and non-stationarity~\cite{coles2001introduction}; our model validation further confirms this (\S\ref{sec:model-eval}). The synchronization tax is governed by the sample maximum $M_n = \max_{i \in [n]} X_i$.

Similar to the Central Limit Theorem (CLT) for sample means, EVT provides limiting distributions for the sample maxima. In our case, each sample is the set of pre-collective GEMM kernel executions across ranks. The Fisher--Tippett--Gnedenko Theorem states that if the maximum of i.i.d. samples converges in distribution, then the limit must be a generalized extreme value distribution (GEVD) \emph{regardless of the underlying kernel runtime distribution}~\cite{coles2001introduction}.

The GEVD is defined by its location ($\mu$), scale ($\sigma > 0$), and shape ($\xi$) parameters. The shape parameter $\xi$ dictates the tail behavior: $\xi > 0$ yields a heavy-tailed Fr\'{e}chet distribution, $\xi < 0$ yields a strictly bounded Weibull, and $\xi \to 0$ yields the light-tailed Gumbel.

\myparab{A conservative estimate.} A distribution converges to the Weibull form of the GEVD when it has a finite upper bound (\eg hard timeout), so values above this bound occur with \emph{strictly} zero probability. Thus, we can safely rule out Weibull because GPU hardware does not impose a strict timeout or upper bound on kernel execution. Between the Fr\'{e}chet and Gumbel distributions, our analysis focuses primarily on the light-tailed Gumbel because it provides the \emph{most conservative} estimate of the synchronization tax over the heavy-tailed Fr\'{e}chet. (The Fr\'{e}chet mean maximum grows as $k^{\xi}$ with $\xi{>}0$ by definition, which dominates the Gumbel's $\ln k$ growth.) However, as we detail below, our empirical data is better fit by the Fr\'{e}chet distribution, so our results provide a \emph{lower} bound on the synchronization tax at scale.

\begin{figure}[h]
\centering
\includegraphics[width=\linewidth]{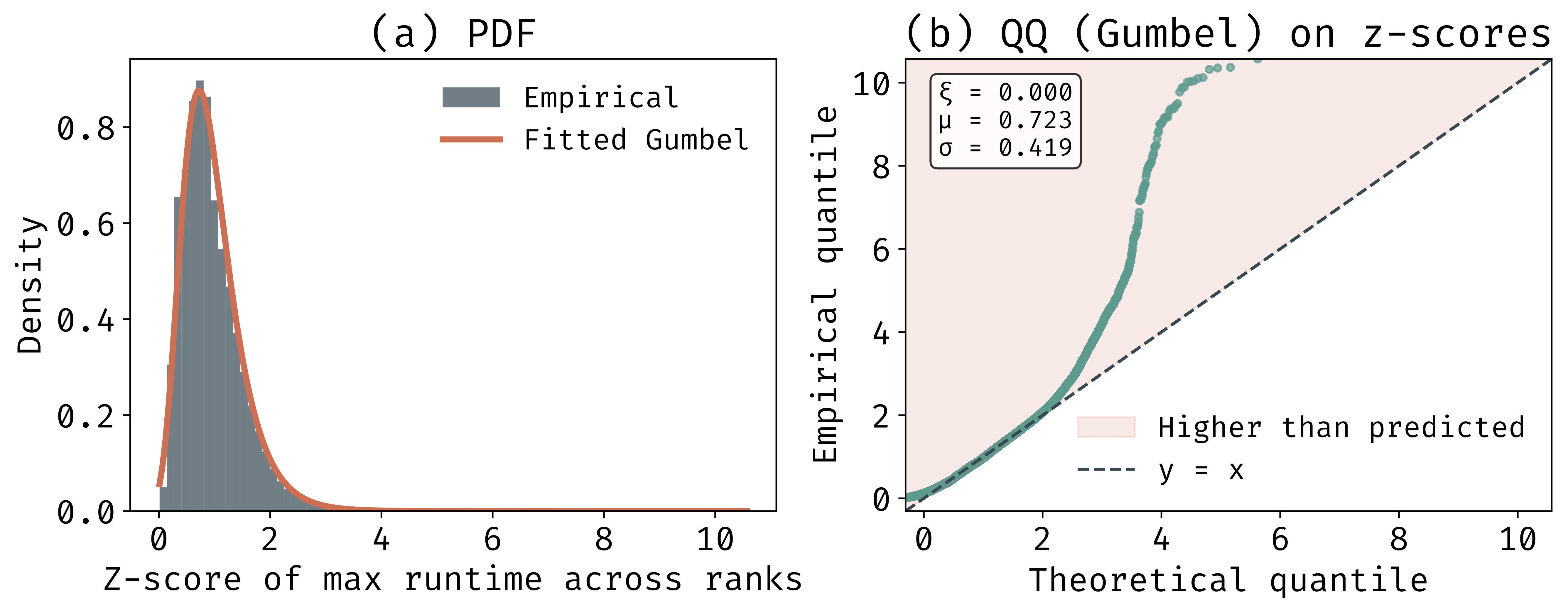}
\caption{Gumbel distribution fitted to $z$-scores of maximum cross-rank compute-block runtimes ($z_{c,j}$) across all workloads.}
\label{fig:gev-fit}
\end{figure}

\myparab{Gumbel fit.} Figure~\ref{fig:gev-fit} shows the Gumbel fit to our data using the standard maximum likelihood estimation (MLE) but fixing the shape parameter as $\xi = 0$ to specify the Gumbel shape. While the probability density function (PDF) fits well (Figure~\ref{fig:gev-fit}, left), the QQ plot (Figure~\ref{fig:gev-fit}, right) suggests that the Gumbel distribution \emph{under-estimates} the synchronization delay in our data, as empirical quantiles for larger $z$-scores fall \emph{above} the fitted distribution's predicted quantiles.

\begin{figure}[h]
\centering
\includegraphics[width=\linewidth]{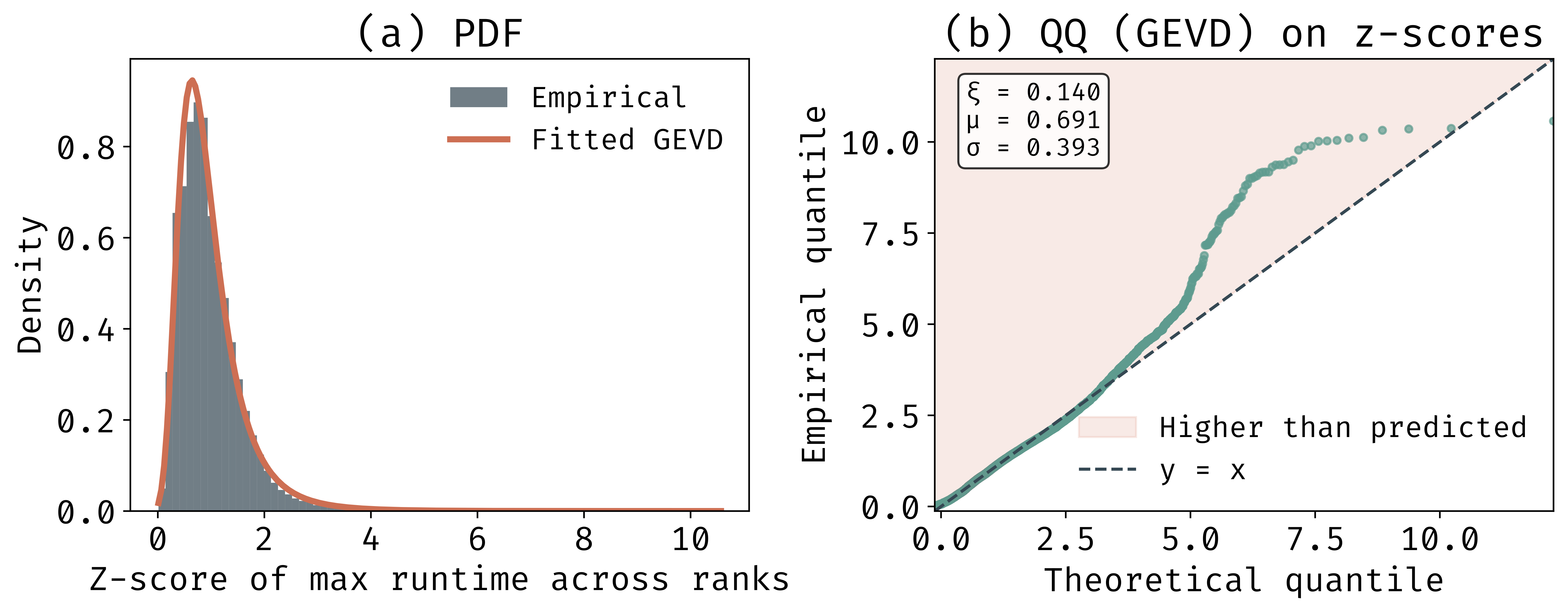}
\caption{Best-fitting GEVD for $z$-scores of the maxima of cross-rank compute-block execution times ($z_{c,j}$) across all hardware architectures and workloads.}
\label{fig:frechet}
\end{figure}

We also fit the GEVD using MLE with \texttt{scipy}, allowing all three parameters to be free (including the shape, $\xi$). Figure~\ref{fig:frechet} shows the best-fitting GEVD to the $z$-scores of per-instance GEMM kernel maxima across all workloads. The best-fitting distribution has a shape parameter of $\xi=0.14$, which falls into the heavy-tailed Frechet domain ($\xi > 0$). Even the fitted distribution shown in Figure~\ref{fig:frechet} may understate the synchronization tax, as empirical quantiles fall above theoretical quantiles, suggesting heavy-tailed behavior. The expected value of a GEVD random variable is
\[
\mathbb{E}[X] = \mu + \sigma \cdot \frac{\Gamma(1 - \xi) - 1}{\xi}, \quad \xi \neq 0, \quad \xi < 1
\]
where $\Gamma(\cdot)$ is the gamma function. When $\xi \geq 1$, the expected value is undefined due to a severely heavy tail.

\myparab{Max stability.} By max-stability, the maximum of $k$ i.i.d. samples from $\mathrm{GEV}(\mu, \sigma, \xi)$ is distributed as $\mathrm{GEV}(\mu + \sigma\frac{k^\xi-1}{\xi},\, \sigma k^\xi,\, \xi)$ \cite{coles2001introduction}. Thus, the expected maximum in a cluster of size $n = km$ is
\begin{equation}
\mathbb{E}[M_n] = \mu + \frac{\sigma}{\xi}\!\left(k^{\xi}\,\Gamma(1-\xi) - 1\right).
\label{eq:gev-ev}
\end{equation}
Unlike the logarithmic growth with Gumbel (Eq.~\ref{eq:gumbel-ev}), the expected maximum here grows as $k^{\xi}$, \ie polynomially in cluster size, implying an even faster-growing synchronization tax. We provide more details about testing for independence and EVT assumptions in \S\ref{sec:fitting}.

\begin{figure}[h]
\centering
\includegraphics[width=\linewidth]{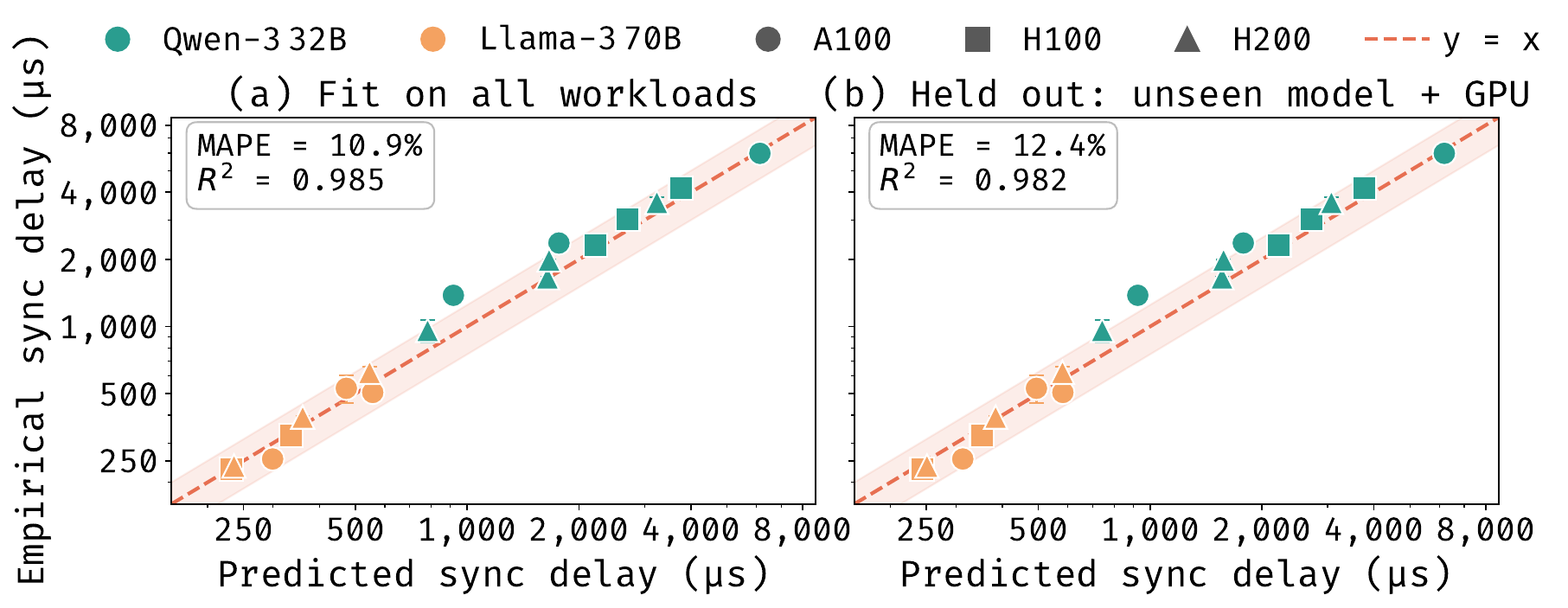}
\caption{Comparison of best-fitting (Frechet) model's prediction to the empirical mean synchronization tax across all collectives following each unique compute block (data point). Error bars are standard error of the mean, and shaded area is $\pm 25\%$.}
\label{fig:gev-validation}
\end{figure}

Using an identical approach as in \S\ref{sec:model-eval}, with both full and leave-one-out validation, we evaluate how well the best-fitting GEVD captures the synchronization tax experienced at the ensuing collective in Figure~\ref{fig:gev-validation}. The accuracy, \ie MAPE, for both the full and leave-one-out validations generally improves from the Gumbel validation in Figure~\ref{fig:gumbel-val}, indicating that the tax is likely heavy-tailed (Frechet) and the Gumbel indeed provides a lower bound.

\begin{figure*}[h]
\centering
\includegraphics[width=0.75\linewidth]{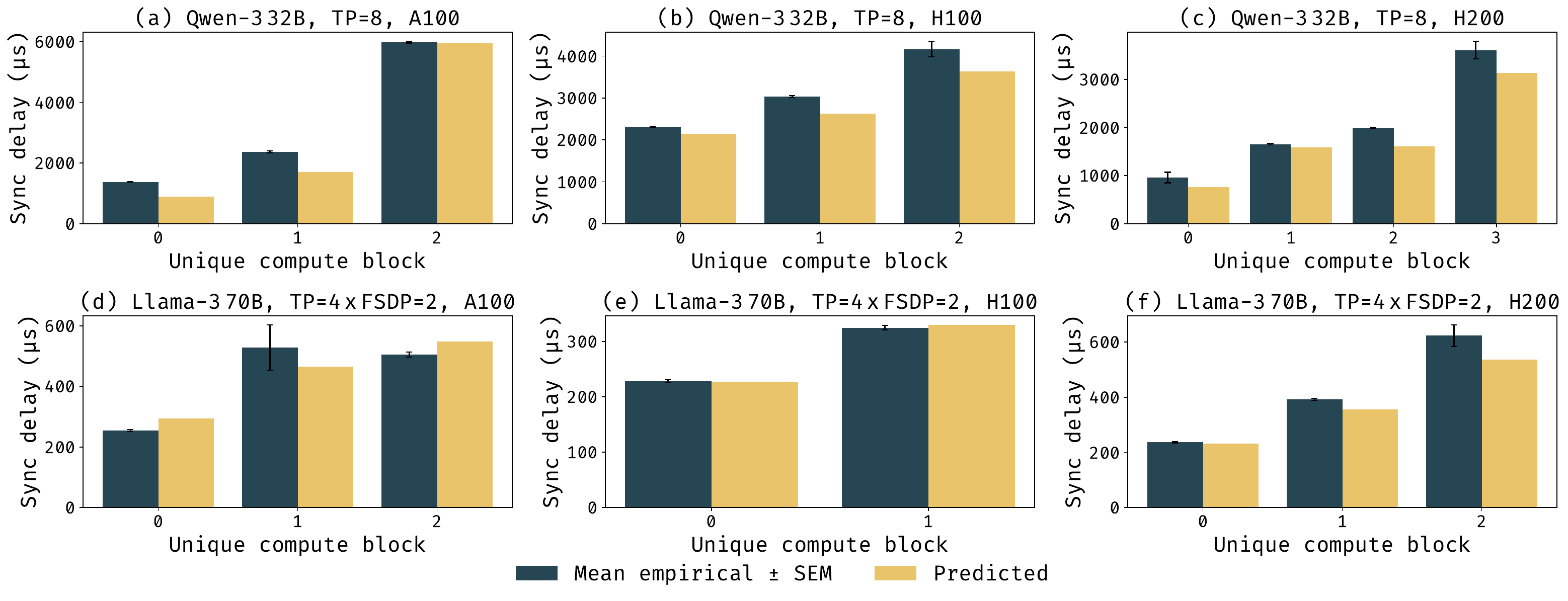}
\caption{Validation of the Gumbel distribution. For each unique compute block (x-axis), we compare the predicted expected delay---derived from the fitted Gumbel distribution and $z$-score conversion---to the empirically observed mean synchronization delay across all collectives following the given compute block. }
\label{fig:gumbel-bars}
\end{figure*}

\begin{figure*}[h]
\centering
\includegraphics[width=0.75\linewidth]{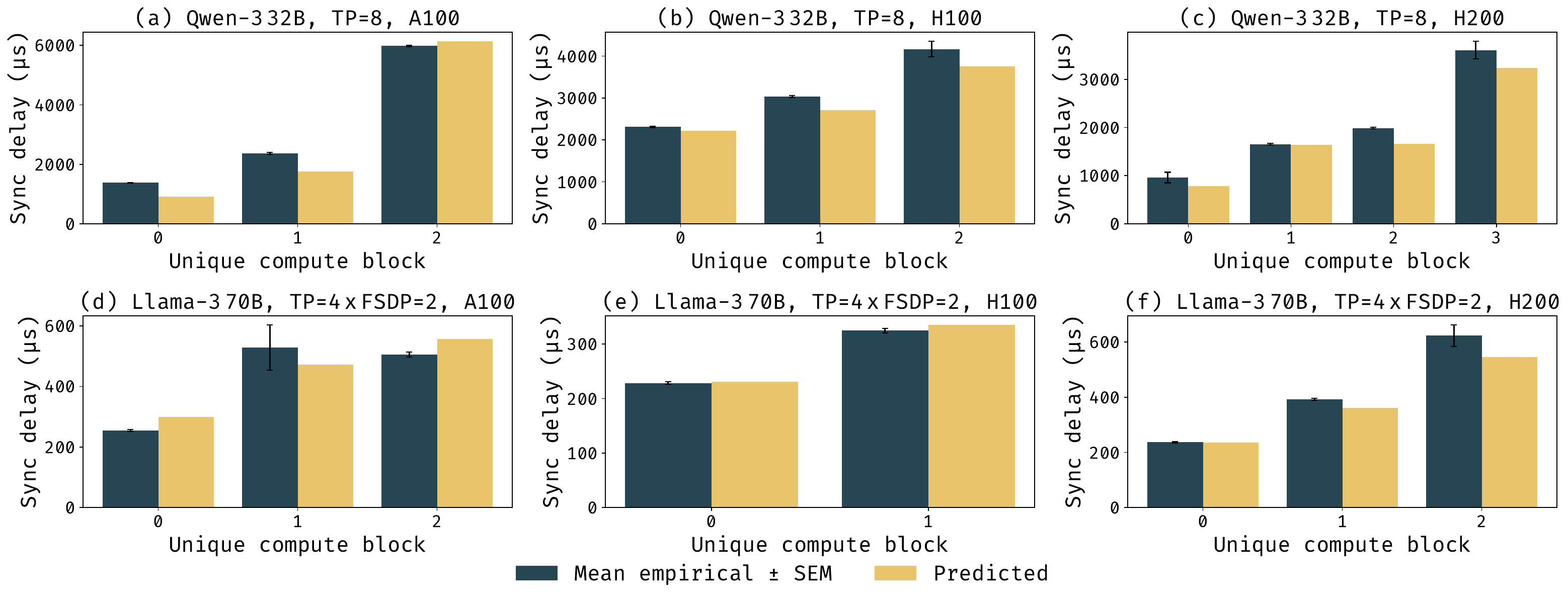}
\caption{Validation of the best-fitting GEV distribution. For each unique compute block (x-axis), we compare the predicted expected delay---derived from the best-fitting GEV distribution and $z$-score conversion---to the empirically observed mean synchronization delay across all collectives following the given compute block. }
\label{fig:gev-bars}
\end{figure*}

\begin{figure*}[h]
\centering
\includegraphics[width=0.8\linewidth]{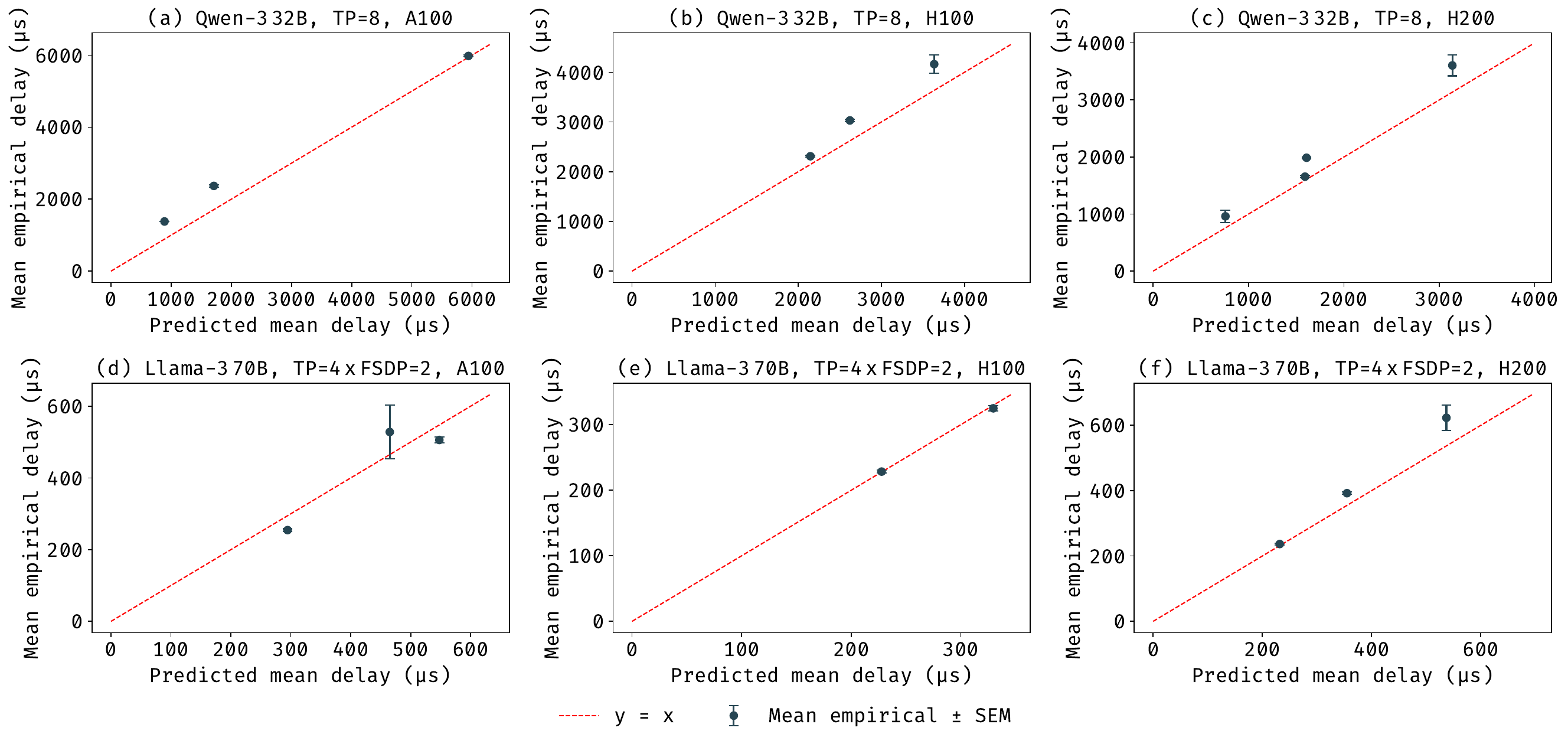}
\caption{Scatterplot of our validation of the Gumbel distribution.}
\label{fig:gumbel-scatter}
\end{figure*}

\begin{figure*}[h]
\centering
\includegraphics[width=0.8\linewidth]{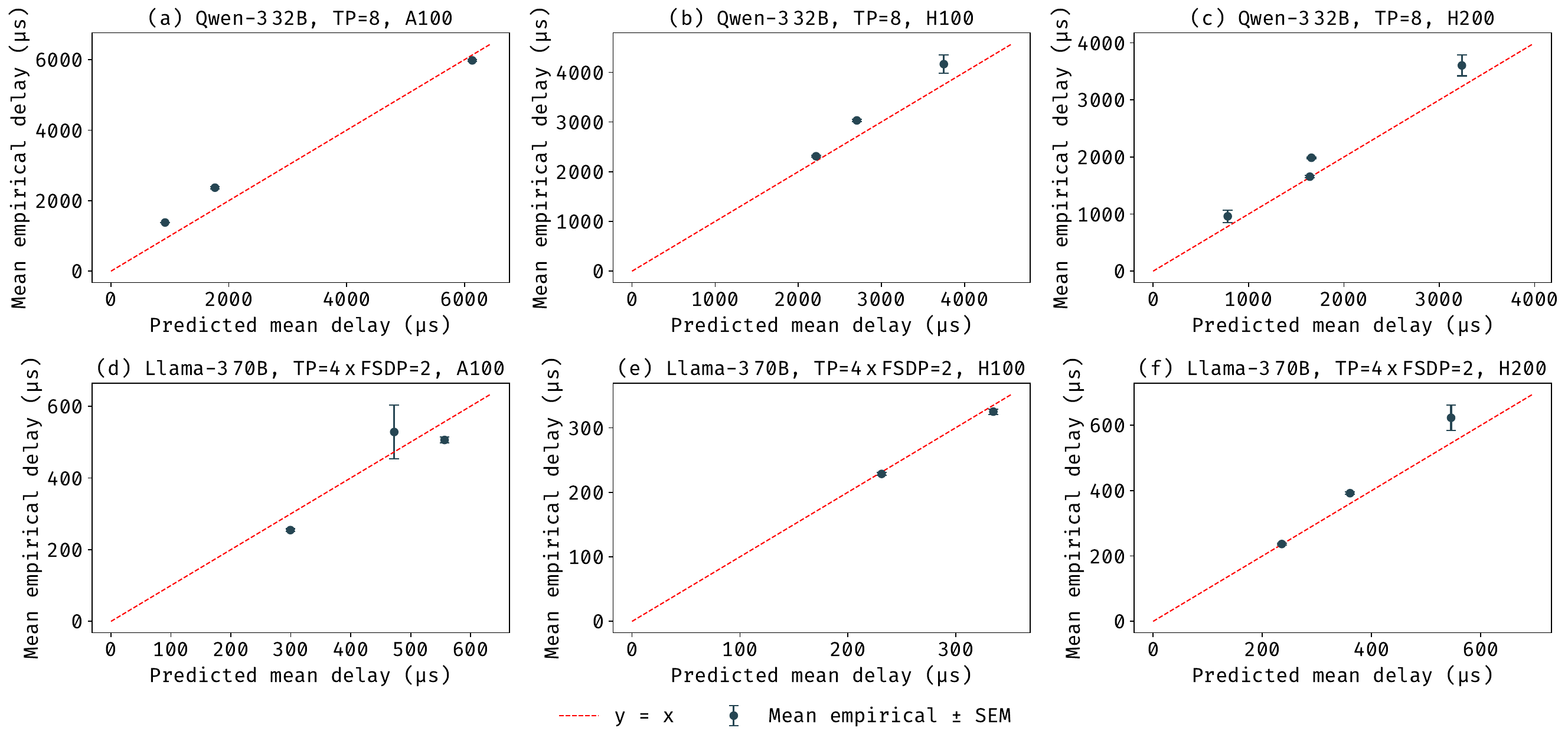}
\caption{Scatterplot of our validation of the best-fitting GEV distribution.}
\label{fig:gev-scatter}
\end{figure*}

We also provide bar graphs (Figure~\ref{fig:gumbel-bars} for Gumbel, Figure~\ref{fig:gev-bars} for best-fit) and  scatterplots (Figure~\ref{fig:gumbel-scatter} for Gumbel, Figure~\ref{fig:gev-scatter} for best-fit) of our model validation for both the Gumbel and best-fitting (Frechet) distributions. (This features validation with our entire dataset, \ie the fits ultimately used for the analytical model in \S\ref{sec:bw-model}, not the leave-one-out validation.) The points falling along the $y{=}x$ line indicate that our analytical model of cross-rank variation in compute block execution can accurately predict the mean observed synchronization delay during the collective.


\section{Calculating the z-score}
\label{sec:pooled-std}
We provide the formulas to calculate the $z$-scores of maximum GEMM compute-block execution times (preceding each collective) across ranks that we fit the GEVD to. In statistics, the pooled standard deviation is used to calculate the combined standard deviation of independent groups from the same population. In our case, the population is all execution times of compute block $c$, and each independent group is a unique instance $j$ of that compute block.  $s_{c,j}$ is the standard deviation across $m$ ranks of instance $j$ of compute-block $c$.
\begin{equation}
\begin{aligned}
M_{c,j} &= \max_{i} X_{c,j,i}, \quad
\bar{X}_{c,j} = \frac{1}{m} \sum_{i=1}^{m} X_{c,j,i}, \\
s_{c,\text{pooled}} &= \sqrt{
\frac{\sum_{j} (m - 1)\, s_{c,j}^{2}}{\sum_{j} (m - 1)}
},
\end{aligned}
\end{equation}

\section{Discussion}
\label{sec:discussion}

\myparab{Is the tax driven by variation during data transfer?} Our algorithm in \S\ref{sec:critical-path} identifies the sequence of events on the critical path \emph{before} each collective. This, by definition, excludes intra-collective operations. Our algorithm accurately traces the synchronization delay at sub-10-$\mu{s}$ granularity to events before the collective, ruling out intra-collective operations as sources of variation. Moreover, unlike the scale-out network~\cite{deng2025mycroft}, scale-up fabrics are designed to be contention-free~\cite{nvlink}, further confirming that variation in compute kernel runtimes drives the synchronization tax.

\myparab{Does the tax extend beyond tensor-parallel (TP) collectives?} The tax for FSDP and EP is even larger than for TP (\S\ref{sec:characterization}, \S\ref{sec:appendix-measurements}). Our analysis focuses on TP to provide a conservative estimate of the synchronization tax, as TP schedules equal work across GPUs, unlike EP, which experiences token imbalance that further exacerbates the tax~\cite{hwang2023tutel}. Further, TP results in frequent ``exposed'' collective communication that is difficult to overlap with computation and is therefore typically restricted to the scale-up domain~\cite{torchtitan,pati2024t3, narayanan2021efficient}. In fact, because of the coupling between parallelisms in the scale-up domain, even in hybrid parallelism the TP group with the slowest rank often sets the domain-wide tax (\S\ref{sec:bw-model}).

\myparab{Is the tax an implementation artifact?} Collective communication libraries, \eg NCCL, provide implementations that guarantee inclusion of every rank's data~\cite{hu2025demystifying}, so a collective cannot complete until \emph{every rank} in the group is ready to communicate, regardless of implementation~\cite{devraj2025efficient}. Even tile-based overlap~\cite{pati2024t3, wang2022overlap,chen2024centauri} and implementations using remote load-store semantics~\cite{devraj2026breaking} still incur the tax, as each tile still executes the GEMM computation that drives the tax. Because we isolate the tax to GEMM kernel variation, it applies in principle to both training and inference and is exacerbated by expert parallelism in MoE models (as shown by the much larger tax in DeepSeek-V3 MoE in \S\ref{sec:characterization}).

\myparab{How about non-collective traffic in the scale-up fabric?} While communication in pre-training and inference primarily consists of collectives, scale fabrics may also carry point-to-point traffic such as KV-cache transfers in disaggregated inference. However, any transfer that depends on \emph{multiple} producer GPUs still pays the synchronization tax because it must wait on this data~\cite{devraj2026breaking}.

\myparab{How do we extrapolate with EVT?} EVT provides a general framework to model maxima without assuming a specific parent distribution~\cite{coles2001introduction, katz2002statistics}, \ie the equivalent of the Central Limit Theorem for maxima. As a result, we are able to design---to our knowledge---the first analytical model of cross-GPU variation under identical compute. Our model is conservative for several reasons. First, we only use EVT to model \emph{fundamental variation} that grows with scale; additional factors, \eg throttling, MoE imbalance, would exacerbate the tax. Second, we constrain our model to use the light-tailed Gumbel distribution, despite our empirical data being better fit by the heavy-tailed Fréchet. We conduct sensitivity analyses, verify key EVT conditions (\S\ref{sec:fitting}), and validate that it predicts synchronization delays in held-out workloads. For Qwen-3 32B, we fit on blocks of size 4 but apply the model to TP groups of 8 ranks, and it accurately predicts both cross-rank compute variation \emph{and} the delay at the collective. We further fit our model on data from just a 4-GPU node but then applying it to a 128-GPU workload, where it accurately predicts the growth of cross-rank variation in GEMM kernels (\S\ref{sec:model-eval}).

\clearpage

\end{document}